\documentclass[12pt,tightenlines,eqsecnum,floats,aps,amsmath,amssymb,nofootinbib,prd,showpacs]{revtex4}

\usepackage{amsmath,amssymb,amsfonts}
\usepackage{graphicx}
\usepackage{enumerate} 
\usepackage{colordvi} 

\def\be{\begin{equation}}
\def\ee{\end{equation}}
\def\ba{\begin{eqnarray}}
\def\ea{\end{eqnarray}}

\def\H{{\cal H}}
\def\Hkg{\H_{\rm kin}^{\rm grav}}
\def\Hk{\H_{\rm kin}^{\rm total}}
\def\Hkwdw{\H_{\rm kin}^{\rm wdw}}

\def\Hp{\H_{\rm phy}}

\def\Hpwdw{\H_{\rm phy}^{\rm wdw}}

\def\V{{\cal V}}
\def\h{\hat }
\def\su{{\rm su}}

\def\Tr{{\rm Tr\,}}

\def\g{\gamma}
\def\lp{{\ell}_{\rm Pl}}
\def\R{\mathbb{R}}
\def\Z{\mathbb{Z}}

\def\q{{}^o\!q}
\def\e{{}^o\!e}
\def\w{{}^o\!\omega}

\newcommand{\ket}[1]{\ensuremath{|#1\rangle}}
\newcommand{\ip}[2]{{\langle#1\,|\,#2\rangle}}
\newcommand{\rcr}{\rho_{\mathrm{crit}}}
\newcommand{\snf}{\sin(\bar \mu c)}
\newcommand{\snfs}{\sin^2(\bar \mu c)}
\newcommand{\csf}{\cos(\bar \mu c)}

\newcommand{\heff}{{\cal H}_{\mathrm{eff}}}

\newcommand{\p}{\partial}

\newcommand{\f}{\frac}

\newcommand{\ep}{\epsilon}

\newcommand{\fracs}[2]{{\scriptstyle\frac{#1}{#2}}} 

\usepackage{enumerate}

\usepackage{colordvi}
\newcounter{mnotecount}[section]

\newcommand{\comment}[1]{}

\def\f{\frac}

\def\La{\mathcal{L}}

\def\dd{\textrm{d}}
\def\ub{\underbar}
\def\ul{\underline}
\def\WDW{WDW\,\,}
\def\t{\tilde}
\def\epsilon{\varepsilon}

\def\hkm{h_k^{(\bar \mu)}}

\def\hkmi{h_k^{(\bar \mu)-1}}

\def\sinm{\sin\left(\f{\mb c}{2}\right)}
\def\cosm{\cos\left(\f{\mb c}{2}\right)}
\def\mb{\bar \mu}

\begin{document}
\preprint{\vbox{\baselineskip=12pt \rightline{IGPG-06/06-}
}}

\title{Quantum Nature of the Big Bang: Improved dynamics}

\author{Abhay Ashtekar${}^{1,2,3}$}
\email{ashtekar@gravity.psu.edu}
\author{Tomasz Pawlowski${}^{1}$}
     \email{pawlowsk@gravity.psu.edu}
    \author{Parampreet Singh${}^{1}$}
    \email{singh@gravity.psu.edu}
\affiliation{${}^{1}$Institute for Gravitational Physics and
Geometry, Physics Department, Penn State, University Park, PA
16802, U.S.A.\\
${}^2$ Institute for Theoretical Physics, University of Utrecht,
Princetonplein5, 3584 CC Utrecht, The Netherlands\\
${}^3$Isaac Newton Institute for Mathematical Sciences, 20
Clarkson Road, Cambridge CB3 0EH, UK}

\begin{abstract}

An improved Hamiltonian constraint operator is introduced in loop
quantum cosmology. Quantum dynamics of the spatially flat,
isotropic model with a massless scalar field is then studied in
detail using analytical and numerical methods. The scalar field
continues to serve as `emergent time', the big bang is again
replaced by a quantum bounce, and quantum evolution remains
deterministic across the deep Planck regime. However, while with
the Hamiltonian constraint used so far in loop quantum cosmology
the quantum bounce can occur even at low matter densities, with
the new Hamiltonian constraint it occurs only at a Planck-scale
density. Thus, the new quantum dynamics retains the attractive
features of current evolutions in loop quantum cosmology but, at
the same time, cures their main weakness.

\end{abstract}

\pacs{04.60.Kz,04.60Pp,98.80Qc,03.65.Sq}

\maketitle

\section{Introduction}
\label{s1}

The spatially flat, isotropic model was recently investigated in
detail in the setting of loop quantum cosmology (LQC) \cite{aps2}.
(For a brief summary of results, see \cite{aps1}.) That
investigation introduced a conceptual framework and analytical and
numerical tools to construct the \emph{physical sector} of the
quantum theory. These methods enabled one to systematically
explore the effects of quantum geometry both on the gravitational
and matter sectors and to extend the previous results in LQC.

If the only matter source is a massless scalar field, \emph{every}
classical solution has a singularity which, furthermore, persists
in the Wheeler-DeWitt theory. Therefore, to bring out effects
which can be unambiguously attributed to the quantum nature of
geometry underlying LQC, this model was analyzed in detail in
\cite{aps1,aps2}. The analysis led to the following results: i)
the scalar field was shown to serve as an internal clock, thereby
providing a detailed realization of the `emergent time' idea; ii)
the physical Hilbert space, Dirac observables and semi-classical
states were constructed rigorously; and, iii) the Hamiltonian
constraint was solved numerically to show that in the backward
evolution of states which are semi-classical at late times,
\emph{the big bang is replaced by a quantum bounce}. Furthermore,
thanks to the non-perturbative, background independent methods,
unlike in other approaches the quantum evolution is
\emph{deterministic across the deep Planck regime}.

These results are attractive and add to the growing evidence
\cite{mb1,abl,mbrev} suggesting that the quantum geometry effects
of loop quantum gravity (LQG) hold a key to many long standing
questions. Control over the physical sector of the theory and
availability of numerical methods also enabled to us to analyze
the physics of the bounce in greater detail than could be done in
previous investigations. These details enriched our understanding
of both the semi-classical and deep Planck regimes. However, they
also brought out a serious limitation of the framework: the
critical value $\rcr$ of the matter density at which the bounce
occurs can be made \emph{arbitrarily small} by increasing the
momentum $p_\phi$ of the scalar field (which is a constant of
motion). Now, large values of $p_\phi$ are permissible and even
preferred for semi-classical considerations at late times. Since
it is physically unreasonable to expect quantum corrections to
significantly modify classical predictions at low matter
densities, this limitation is a serious drawback. Indications of
such problems had appeared also in some of the earlier results
based on effective equations (see, e.g., \cite{dh2,mbrev}).
However, since the approximations inherent to effective equations
break down in the deep Planck regime, one did not have an a priori
reason to believe that these were consequences of the underlying
\emph{quantum} equations of LQC, rather than artifacts of
approximation schemes. The analysis of \cite{aps2} unambiguously
showed that the problem lies with the quantum equations
themselves. Thus, while the quantum evolution predicted by the
existing LQC constraint operator has a number of attractive
features, it has one major weakness. A key question then is
whether one can change the definition of the constraint operator
in a subtle way so that this problem is overcome, but the
attractive features of the evolution are all preserved.

As reported in \cite{aps2,aps1}, the answer is in the affirmative.
The purpose of this paper is to justify this assertion by
providing the detailed construction of the new Hamiltonian
constraint and analyzing the resulting quantum dynamics. We will
find that the required change is in fact physically well motivated
and conceptually quite compelling.

The main ideas can be summarized as follows. Recall that, to
obtain the expression of the quantum constraint, one has to first
introduce an operator representing curvature of the gravitational
connection. Now, a key feature of LQC ---inherited directly from
full loop quantum gravity (LQG)--- is that while there are well
defined quantum analogs of holonomies, there is no operator
corresponding to the connection itself \cite{alrev,crbook,ttbook}.
Thus, one is naturally led to define the curvature operator in
terms of holonomies. In the classical theory, curvature can be
expressed as a limit of the holonomies around a loop as the area
enclosed by the loop shrinks to zero. In quantum geometry,
however, one can not continuously shrink the loop to zero area;
there is a smallest non-zero area eigenvalue, or an `area gap'
$\Delta$ \cite{rs,almmt,al5}. The physical idea then is to
incorporate the existence of the mass gap $\Delta$ \cite{abl} in
the limiting procedure. However, in the existing definition of the
quantum constraint, this physical idea was mathematically
implemented in a rather indirect manner. Initially, it was meant
to be only a first step to gain a qualitative understanding of
dynamics within LQC. Nonetheless, as more and more physically
appealing features of this dynamics were discovered, the
constraint operator obtained by this procedure was taken more and
more seriously and constituted the basis of a number of
approximation schemes and effective descriptions.

The new Hamiltonian constraint introduced in this paper is based
on the same principles but the curvature operator is now
constructed by a more direct implementation of the underlying
physical ideas. The structure of the final constraint operator
does change significantly but, when suitably recast, its form is
actually somewhat simpler. Furthermore, it has additional
conceptually attractive features. For example, its Wheeler-DeWitt
(WDW) limit naturally comes with the `covariant factor ordering'
of that theory. Once the new constraint is introduced, we apply
analytical and numerical methods of \cite{aps2} to extract
physical predictions from the theory. Specifically, we will build
the physical Hilbert space $\Hp$ from solutions to the new quantum
constraint, introduce on it a complete family of Dirac
observables, construct suitable families of states which are
semi-classical at `late times' (e.g., `now'), and numerically
evolve them back in time. As in \cite{aps2}, we will find that in
this backward evolution, the big bang is again replaced by the
quantum bounce. Qualitative features of the evolution are very
similar to those observed in \cite{aps2}, \emph{except that the
big bounce occurs precisely when the matter density enters the
Planck regime,} irrespective of the value of $p_\phi$ and other
initial conditions.

To summarize, the Hamiltonian constraint will be improved by
implementing the physical idea of \cite{abl} in a more
satisfactory manner. The resulting dynamics will retain the
attractive features of the older quantum evolution but free it
from its main drawback. This rather delicate interplay between
physics and mathematics is both interesting and instructive. We
wish to emphasize however that the present procedure still retains
a basic limitation of the existing treatments: the Hamiltonian
constraint is not systematically derived from the full theory.
Indeed, this important task can not yet be undertaken because
there is no unambiguous Hamiltonian constraint in the full theory
which can serve as the natural starting point for a systematic
reduction.

The organization of this paper parallels that of \cite{aps2} and
we use the same notation. \emph{We direct the reader to that paper
for motivations as well as further technical details.} In Section \ref{s2}
we introduce the new Hamiltonian constraint operator. In Section
\ref{s3} we discuss the \WDW theory that results when one ignores
the effects of quantum geometry. As in \cite{aps2}, we find that
the singularity is not resolved in that limit. In Section \ref{s4}
we return to LQC and construct the physical sector of the theory.
Quantum states which are semi-classical at `late times' are then
numerically evolved backwards in Section \ref{s5}. We find that
the classical big bang is replaced by a quantum bounce which
occurs when the matter is compressed enough to acquire a density
of the Planck scale. Thus, in the deep Planck regime, quantum
geometry has the effect of making gravity strongly repulsive. A
key virtue of the new constraint operator is that these effects
are completely negligible when the matter density is significantly
smaller than the Planck density. Section \ref{s6} summarizes the
main results and briefly discusses possible extensions. While our
detailed analysis is restricted to a specific and rather simple
model, the key ideas behind the construction of the new constraint
operator can be applied much more generally. This is illustrated
in Appendix \ref{a1} where we allow for the presence of a
cosmological constant.  (Other models will be discussed also by
other authors in forthcoming papers.) Appendix \ref{a2} discusses
conceptual issues that are especially relevant to approximation
schemes used in constructing effective equations.

A brief overview of LQG emphasizing these developments in LQC can
be found in \cite{aaparis}.

\section{The improved Constraint Operator}
\label{s2}

This section is divided into three parts. In the first two we
introduce the key new elements and in the third we put them
together to construct the new Hamiltonian constraint operator.

\subsection{Strategy}
\label{s2.1}

Let us first collect the required background material and
notation. (For details, see \cite{abl,aps2}.) In a systematic
Hamiltonian treatment of spatially flat, isotropic models, one has
to first introduce an elementary cell $\V$ and restricts all
integrations to this cell. Fix a fiducial flat metric $\q_{ab}$
and denote by $V_o$ the volume of the elementary cell $\V$ in this
geometry. The gravitational phase space variables ---the
connections $A_a^i$ and the density weighted triads $E^a_i$--- can
be expressed as:
\be A_a^i = c\,V_o^{-\f{1}{3}} \w_a^i, \quad {\rm and} \quad E^a_i
= p\, V_o^{-\f{2}{3}} \sqrt{q_o}\,\, \e^a_i \ee
where $(\w_a^i, \e^a_i)$ are a set of orthonormal co-triads and
triads compatible with $\q_{ab}$ and adapted to the edges of the
elementary cell $\V$. Thus, the symmetry reduced gravitational
phase space is only 2-dimensional, coordinatized by the pair $(c,
p)$. Thanks to the availability of the fiducial cell $\V$, the
coordinates $(c,p)$ are insensitive to the choice of the fiducial
metric, i.e., remain unchanged under the rescaling $\q_{ab}
\rightarrow \alpha^2 \q_{ab}$.  The fundamental Poisson bracket is
given by:
\be \{c,\, p\} = \f{8\pi G\g}{3} \ee
where $\g$ is the Barbero-Immirzi parameter. The gravitational
part of the Hamiltonian constraint can be written as:
\be C_{\mathrm{grav}} = - \gamma^{-2} \int_{\cal V} d^3 x\,
\epsilon_{ijk} \, e^{-1} E^{ai}E^{bj}\, \, F_{ab}^k =
-\f{6}{\g^2}\, c^2 \sqrt{p} \ee
where $e := \sqrt{|\mathrm{det}E|}$.

In quantum theory, following Dirac, one first constructs a
kinematical description. The Hilbert space $\Hkg$ is the space
$L^2(\R_{\rm Bohr}, d\mu_{\rm Bohr})$ of square integrable
functions on the Bohr compactification of the real line. To
specify states concretely, it is convenient to work with the
representation in which the operator $\hat{p}$ is diagonal.
Eigenstates of $\hat{p}$ are labelled by a real number $\mu$ and
satisfy the orthonormality relation:
 \be \ip{\mu_1}{\mu_2} = \delta_{{\mu_1},\, {\mu_2}}\, . \ee
Since the right side is the Kronecker delta rather than the Dirac
delta distribution, a typical state in $\Hkg$ can be expressed as
a countable sum; $\ket\Psi = \sum_n c^{(n)} \ket{\mu_n}$ where
$c^{(n)}$ are complex coefficients and the inner product is given
by
\be \langle \Psi_1| \Psi_2 \rangle = \sum_n \, \bar c_1^{(n)} \,
c_2^{(n)} ~. \ee
%
The fundamental operators are $\hat{p}$ and $\widehat{\exp
i\lambda(c/2)}$:
\be \hat{p}\,\ket{\mu} = \f{8\pi \g \lp^2}{6}\, \mu\, \ket{\mu}
\quad {\rm and} \quad \widehat{\exp {\f{i\lambda
c}{2}}}\,\ket{\mu} = \ket{\mu+\lambda}\ee
where $\lambda$ is any real number. Since the holonomy
$h_k^{(\lambda)}$ of the gravitational connection $A_a^i$ along a
line segment $\lambda\, \e^a_k$  is given by:%
\footnote{Here $\mathbb{I}$ is the unit $2\times 2$ matrix and
$\tau_k$ is a basis in the Lie algebra $\su(2)$ satisfying
$\tau_i\, \tau_j = \f{1}{2}\epsilon_{ijk}\tau^k - \f{1}{4}
\delta_{ij}$. Thus, $2i \tau_k = \sigma_k$, where $\sigma_i$ are
the Pauli matrices.}
\be \label{hol} {h^{(\lambda)}_k} = \cos \fracs{\lambda
c}{2}\,\mathbb{I} + 2 \sin \fracs{\lambda c}{2}\, \tau_k \ee
the corresponding holonomy operator has the action:
\be \h{h}^{(\lambda)}_k\,\ket{\mu} = \f{1}{2}\, \left(
\ket{\mu+\lambda} + \ket{\mu -\lambda} \right) \mathbb{I} +
\f{1}{i}\, \left( \ket{\mu+\lambda} - \ket{\mu -\lambda} \right)
\tau_k  ~.\ee
However, just as there is no operator corresponding to the
connection itself in full LQG \cite{almmt,alrev,crbook,ttbook},
there is no operator $\hat{c}$ on $\Hkg$ \cite{afw,abl}.

To describe quantum dynamics, we have to first introduce a
well-defined operator on $\Hkg$ representing the Hamiltonian
constraint $C_{\rm grav}$. Since there is no operator
corresponding to $c$ itself, we return to the integral expression
of the constraint from the full LQG:
\be \label{eq:cgrav} C_{\mathrm{grav}} = - \gamma^{-2} \int_{\cal
V} d^3 x\,  \epsilon_{ijk} \, e^{-1} E^{ai}E^{bj}\, F_{ab}^i ~.
\ee
For passage to quantum theory, we first need to express this
classical constraint in terms of the elementary variables $p$ and
$h_k^{(\lambda)}$ which have unambiguous quantum analogs. As in
the full theory \cite{tt,ttbook}, the term involving triads can be
written as
\be \label{cotriad} \epsilon_{ijk}\, e^{-1}\,E^{aj}E^{bk}\, =\,
\sum_k \f{({\rm sgn}\,p)}{2\pi\gamma G\lambda\, V_o^{\f{1}{3}}}\,
\, {}^o\!\epsilon^{abc}\,\, \w^k_c\,\, \Tr\left(h_k^{(\lambda)}\,
\{h_k^{(\lambda)}{}^{-1}, V\}\, \tau_i\right) \ee
where $V= |p|^{3/2}$ is the volume of elementary cell $\V$ in the
physical metric determined by $p$ and the holonomy
$h_k^{(\lambda)}$ is evaluated along the segment $\lambda\,
\e^a_k$ (i.e.,  a segment of oriented length $\lambda$ along the
$k$th edge of the elementary cell $\V$). Note that this identity
holds for any choice $\lambda$, even when it is allowed to be a
function of $p$. Let us allow for this possibility and fix the
appropriate $\lambda$ at the end.

For the field strength $F_{ab}^i$, we use the standard strategy
employed in gauge theories. Consider a square $\Box_{ij}$ in the
$i$-$j$ plane spanned by a face of the elementary cell, each of
whose sides has length $\lambda V_o^{1/3}$ with respect to the
fiducial metric $\q_{ab}$. Then, `the $ab$ component' of the
curvature is given by:
\be \label{F} F_{ab}^k\, = \, -2\,\lim_{Ar_\Box
  \rightarrow 0} \,\, \Tr\,
\left(\f{h^{(\lambda)}_{\Box_{ij}}-1 }{\lambda^2V_o^{2/3}} \right)
\,\, \tau^k\, \w^i_a\,\, \w^j_b\, . \ee
Here $Ar_\Box$ is the area of the square under consideration, and
the holonomy $h^{(\mu_o)}_{\Box_{ij}}$ around the square
$\Box_{ij}$ is just the product of holonomies (\ref{hol}) along
the four edges of $\Box_{ij}$:
\be
 h^{(\lambda)}_{\Box_{ij}}=h_i^{(\lambda)} h_j^{(\lambda)}
 (h_i^{(\lambda)})^{-1} (h_j^{(\lambda)})^{-1}\, . \ee
Combining Eqs. (\ref{cotriad}) and (\ref{F}), $C_{\mathrm{grav}}$
can be written as
\ba \label{hc1} C_{\mathrm{grav}} &=& \nonumber \lim_{Ar_\square
\rightarrow 0} C_{\mathrm{grav}}^{(\lambda)}\quad \mathrm{where} \\
C_{\mathrm{grav}}^{(\lambda)} &=& -\f{4 \,\mathrm{sgn}(p)}{8 \pi
\gamma^3 \lambda^3 G} \, \sum_{ijk} \,\epsilon^{ijk} \,
\mathrm{Tr}\left(h_i^{(\lambda)} h_j^{(\lambda)}
(h_i^{(\lambda)})^{-1} (h_j^{(\lambda)})^{-1} h_k^{(\lambda)}
\{(h_k^{(\lambda)})^{-1},\, V \} \right)\nonumber\\
&=&  \sin\lambda c \left[-\, \f{4}{8\pi G\g^3} \f{{\rm
sgn}(p)}{\lambda^3}\, \sum_k \Tr \tau_k h_k^{(\lambda)}\,
\{(h_k^{(\lambda)})^{-1},\, V\}\right] \sin \lambda c  \ea
where in the last step we have used a symmetric ordering of the
three terms for later convenience. Since the constraint is now
expressed in terms of elementary variables which have unambiguous
quantum analogs, it is straightforward to write down the quantum
operator $\h{C}_{\mathrm{grav}}^{(\lambda)}$. However, the limit
$Ar_\Box \rightarrow 0$ of this operator does not exist. This is
not accidental; had the limit existed, there would be a
well-defined local operator corresponding to the curvature
$F_{ab}^i$ and we know that even in full LQG, while holonomy
operators are well defined, there are no local operators
representing connections and curvatures. This feature is
intimately intertwined with the quantum nature of Riemannian
geometry of LQG. The viewpoint in LQC is that the failure of the
limit to exist is a reminder that there is an underlying
\emph{quantum} geometry where eigenvalues of the area operator are
\emph{discrete}, whence there is a smallest non-zero eigenvalue,
$\Delta$, i.e., an \emph{area gap}. Thus, quantum geometry is
telling us that, on physical grounds, it is inappropriate to let
$Ar_\Box$ go to zero; the `correct' procedure must take into
account the existence of the area gap.

Up to this stage, we have followed the standard LQC reasoning,
first introduced in \cite{abl}. The difference comes in the
implementation of the above idea. The viewpoint we now adopt is
that since quantization of area refers to \emph{physical}
geometries, we should shrink the loop $\Box_{ij}$ till the area
enclosed by it, as measured by the physical metric $q_{ab}$,
reaches the value $\Delta$. Since the physical area of faces of
the elementary cell is $|p|$ and since each side of $\Box_{ij}$ is
$\lambda$ times the edge of the elementary cell, we are led to
choose for $\lambda$ a specific function $\mb(p)$, given by
\be \label{mbar} \mb^2\,\, |p|\, =\, \Delta\, \equiv \,
(2\sqrt{3}\pi\g) \, \lp^2 \ee

While $\mb$ is a non-trivial function on the phase space, its
analog $\mu_o$ in the existing LQC treatments was required to
be a constant.%
\footnote{The procedure used in LQC (and specifically in
\cite{abl,aps2}) so far can be summarized as follows. One notes
that, regarded as a state in the connection representation, each
holonomy $h_k^{(\mu_o)}$ is an eigenstate of the area operator
(associated with the face of the elementary cell orthogonal to the
$k$th direction). One fixes $\mu_o$ by demanding that this
eigenvalue be $\Delta$ and finds $\mu_o = 3\sqrt{3}/2$, a
constant.}
Technically, this contrast turns out to be important. If we write
the quantum constraint using the $\ket\mu$ basis, the constraint
operator of \cite{abl,aps2} is a difference operator with
\emph{uniform} step size, given by $4\mu_o$. In this basis the
action of the new quantum constraint would be again given by a
difference operator, but the step size now depends on the state
$\ket\mu$ it operates on. This difference turns out to be subtle
enough to remove the major weakness of the current LQC constraint
operator, while retaining its physically desirable features.
Conceptually, the new strategy  has important ramifications for
the fundamental curvature operator. In both treatments, it is
\emph{non-local}. However, while the non-local operator used in
the existing literature depends only on the connection, the
non-local operator introduced here depends both on the connection
\emph{and} the geometry. In the classical limit, effects of
quantum geometry are negligible and the classical limits of both
operators yield the standard expression of the curvature $F_{ab}$.

Finally, some heuristics involving full LQG can be used rather
effectively to bring out the physical difference between the two
strategies. In the sector of full LQG consisting of a small but a
finite neighborhood of homogeneous isotropic cosmologies, one
should be able to carry out partial gauge fixing, enabling one to
speak of 3-metrics (rather than their equivalence classes under
diffeomorphisms). Let us consider quantum geometry states based on
graphs and for simplicity focus on the fixed fiducial cell. Then,
one would expect that, as the scale factor $a$ grows, the number
$N$ of vertices in the graph contained in the fiducial cell would
grow as $N = K_o a^3$ for some constant $K_o$. In the Hamiltonian
constraint of full LQG, the operator $F_{ab}$ would have to be
evaluated at these vertices. A natural procedure is to introduce
an `elementary cube' around each vertex and calculate the holonomy
around the faces of that cube. Since the number $N$ of elementary
cubes contained in the fiducial cell increases as $a$ grows, the
area of their faces, \emph{as measured by the fixed fiducial
metric} would decrease as $Ar_{\rm ele}^o\sim (1/N)^{2/3} \sim
1/|p| $. Hence, as measured by the fiducial metric, the length of
each edge of the face around which the holonomy is evaluated would
go as $1/\sqrt{|p|}$ --- i.e. precisely as $\bar\mu$. Thus the
functional dependence of $\bar\mu$ on $\mu$ can be thought of as
the remnant left behind by the sector of the full theory of
interest to cosmology, on the mini-superspace considered in this
paper. Fixing the edge length to be $\mu_o$, i.e., making it state
independent, amounts to ignoring the `creation of new vertices'
that one expects to accompany the expansion of the universe. These
considerations complement ideas recently advocated by Bojowald to
incorporate inhomogeneities in LQC \cite{boj-pc}.

\subsection{The operator $\widehat{\exp i (\mb c/2)}$}
\label{s2.2}

Having fixed $\lambda$ to be the specific function $\mb(p)$, we
are now led to write the expression of the quantum constraint by
replacing $\lambda$ in the right side of (\ref{hc1}) by $\mb$, the
holonomies and the volume functions by corresponding operators,
and the Poisson bracket by $1/i\hbar$ times the commutator.
However, to carry out this task, we first need to define the
operators $\widehat{\cos (\mb c/2)}$ and $\widehat{\sin (\mb c/2)}$,
i.e. the operator $\widehat{\exp i(\mb c/2)}$. This task is a
somewhat subtle because i) there is no operator corresponding to
$c$ and hence to $\mb c$, and, ii) since $\mb$ is a function of
$p$, ${\exp i(\mb c/2)}$ can not be expressed as a function of the
`elementary variables'  ---$p$ and $\exp i (k c/2)$ where $k$ is a
constant.

However, in this task we are aided by geometric considerations.
Let us begin with states $\t\Psi(\mu)$ in the Schr\"odinger Hilbert
space, $L^2(\R, d\mu)$. For any real constant $k$ we have:
\be \label{holop}\widehat{e^{i \f{k c}{2}}}\, \t\Psi (\mu)\, =\,
\t\Psi(\mu +k ) \,\equiv\,  e^{k\f{\dd}{\dd \mu}}\, \t\Psi (\mu)\,
. \ee
Thus, the action of the operator is to drag the state along the
vector field $k\f{\dd}{\dd \mu}$. Now, the 1-parameter family of
diffeomorphisms generated by the vector field $\dd /\dd \mu$ has a
well-defined action also on the Hilbert space $\Hkg$ now under
consideration. Therefore, Eq (\ref{holop}) holds also for states
$\t\Psi(\mu)$ in $\Hkg$ provided $\exp k\dd/\dd \mu$ is
interpreted simply as the operator that drags the state a unit
affine parameter distance along the vector field $k\f{\dd}{\dd
\mu}$. This suggests that it is natural to set
\be \widehat{e^{i \f{\mb c}{2}}}\, \t\Psi (\mu) =
e^{\mb\f{\dd}{\dd \mu}}\, \t\Psi (\mu) \ee
so the right side would be the image of $\t\Psi$ under the finite
diffeomorphism obtained by moving a unit affine parameter
distance along the integral curve of the vector field $\xi =
\mb\f{\dd}{\dd \mu}$. The affine parameter $v$ along this vector
field is given by:
\be \label{v} v = K \,{\rm sgn}(\mu)\, |\mu|^{\f{3}{2}}, \quad
{\rm where} \quad K =  \f{2\sqrt{2}}{3\sqrt{3\sqrt{3}}} \ee
Although the geometrical meaning of this action of $\widehat{\exp
i (\mb c/2)}$ is simple, since $v$ is a rather complicated
function of $\mu$, its expression in the $\mu$-representation is
complicated:
\be \label{mbop1} \widehat{e^{i \f{\mb c}{2}}}\, \t\Psi(\mu) =
\t\Psi({\rm sgn}(\t\mu) \mid \t\mu\mid^{\f{2}{3}}),\quad {\rm
where}\quad \t\mu = {\rm sgn}(\mu)\, |\mu|^{\f{3}{2}} +\f{1}{K}\,
\ee
However, it is well-defined because $v(\mu)$ is an invertible (and
$C^1$) function of $\mu$. Furthermore, for $\mu \gg 1$, the action
reduces to the form
\be \widehat{e^{i \f{\mb c}{2}}}\, \t\Psi(\mu) \approx
\t\Psi(\mu+\mb)\ee
familiar from the `standard' LQG analysis where $\mb$ is replaced
by a constant $\mu_o$ (see Eq (\ref{holop})). However, as is clear
from the exact expression (\ref{mbop1}), this form with a `simple
displacement' of the argument is highly inaccurate for small
$\mu$.

The complicated action of $\widehat{\exp i (\mb c/2)}$ just
reflects the fact that the variable $\mu$ determined by
eigenvalues of $\hat{p}$ is not well-adapted to the vector field
$\xi$. Let us therefore change the basis $\ket\mu$ to $\ket{v}$.
This basis is more directly adapted to the volume operator
$\hat{V}$ (associated with the elementary cell $\V$):
\be \label{Vv} \h{V}\ket{v} =  (\f{8\pi\g}{6})^{\f{3}{2}}\, \f{|v|}{K}\,\lp^3
\,\ket{v} \ee
(It follows from (\ref{v}) that, like $\mu$, $v$ is
dimensionless.) These kets also constitute an orthonormal basis in
$\Hkg$: $\ip{v_1}{v_2} = \delta_{v_1,\, v_2}$. The two bases are
of course closely related because $\hat{V}$ and $\hat{p}$ are just
functions of one another. In terms of representations, we can
simply set $\Psi(v) = \t\Psi(\mu)$. In the $v$-representation, the
action of $\widehat{\exp i (\mb c/2)}$ is extremely simple :
\be \label{mbop2} \widehat{e^{i \f{\mb c}{2}}}\, \Psi(v) =
\Psi(v+1) \ee
Therefore, in what follows we will use the $v$ representation.
Many of the expressions obtained using the new constraint will
then have a form rather similar to those obtained in
\cite{abl,aps2}. But it is important to keep in mind that while
those expressions were written in the $\mu$ representation, the
ones in this paper are written in the $v$-representation. This
difference is a key reason why the present Hamiltonian constraint
is free of the drawback of the one used in the literature so far.

Note that the operator $\widehat{\exp i (\mb c/2)}$ is unitary. We
can extend the definition in the obvious manner: By setting
\be \widehat{e^{i k\f{\mb c}{2}}}\, \Psi(v) = \Psi(v+k) \ee
for any constant $k$, we obtain a unitary representation on $\Hkg$
of the 1-parameter group of diffeomorphisms generated by the
vector field $\xi$. On the classical phase space, the function
$\mb c/2$ generates a 1-parameter family of (finite) canonical
transformations. It is precisely the lift to the phase space of
the 1-parameter group of diffeomorphisms on the $p$ or $\mu$
space, generated by the vector field $\mb \dd /\dd \mu$. The
action of $\widehat{\exp i (\mb c/2)}$ simply promotes this
relation to the quantum theory.

\textbf{Remark:} While the geometric interpretation of the action
of $\widehat{\exp i (\mb c/2)}$ makes its definition quite
natural, one might nonetheless ask for the relation between our
factor ordering and that normally used. We will now argue that the
choice we have specified in fact originates in ordinary quantum
mechanics.

Consider 1-d Schr\"odinger quantum mechanics, where states
$\t\Psi(x)$ are generally taken to be functions in $L^2(\R, \dd
x)$. Consider a function $f(x)p$ on the classical phase space. The
canonical transformation it generates is again the lift to the
phase space of the diffeomorphism generated by the vector field
$\xi = f(x) \f{\dd}{\dd x} $ on the configuration space (i.e., the
x-axis). One would expect that this geometric action would be
carried over also to quantum theory. Is this in fact the case? In
the textbook treatments one does not take into account such
geometric considerations but constructs a self-adjoint operator
$\widehat{(fp)}:= (\hbar/2i)\, (f\,\f{\dd}{\dd x} + \f{\dd}{\dd
x}\,f )$ using a symmetric ordering. Its action
$\widehat{(fp)}\t\Psi = (\hbar/i)\, (f\,\f{\dd}{\dd x}\t\Psi
+\f{1}{2}\,\f{\dd}{\dd x}\, f \, \t\Psi)$ does \emph{not}
correspond to the Lie-derivative of $\t\Psi$ with respect to the vector
field $\xi$, which is captured just in the first term.  (The
second term, $\f{1}{2}\,\f{\dd}{\dd x}\, f \, \t\Psi$, multiplies
$\t\Psi$ by the divergence of the vector field $\xi$, i.e., by the
Lie-derivative of the Lebesgue measure with respect to $\xi$.)
Hence, the action of the 1-parameter group of unitary
transformations $\h{U}(\lambda) := \exp i \widehat{(fp)}$ on wave
functions $\t\Psi(x)$ is \emph{different} from that of the
diffeomorphisms generated by $\xi$. However, this is simply
because the Lebesgue measure $\dd x$ fails to be invariant under
the diffeomorphism.

Let us represent quantum states by densities $\Psi$ of weight 1/2
---or, more precisely, half-forms--- so that the scalar product is
given simply by $(\Psi_1, \Psi_2) = \int_\R \bar\Psi_1\,\Psi_2$
without the need of a measure. (Thus, $\bar{\t\Psi}_1\t\Psi_2 \dd
x \sim \bar\Psi_1\Psi_2$. For details, see, e.g., \cite{nmjw}.) On
these half forms, the action of $\widehat{(fp)}$ is
\emph{precisely} that of a Lie derivative; $\widehat{(fp)}\, \Psi
= (\hbar/i)\, {\cal L}_\xi \Psi$. Therefore, the action of the
unitary transformation $U(\lambda)$ is given by:
$$ \h{U}(\lambda) \Psi(v)\, =\, \Psi(v+\lambda\hbar) $$
where $v$ is the affine parameter of the vector field $\xi$ on
$\R$. Thus, the initial geometric expectation is indeed realized
if one proceeds with the standard factor ordering from
Schr\"odinger quantum mechanics, but represents states by half
forms. In the polymer representation the measure is invariant
under the action of any diffeomorphism on the $\mu$-axis. Hence
the additional `divergence term' is unnecessary. In this sense,
states in $\Hkg$ are analogs of densities of weight 1/2 ---or,
half forms--- in the Schr\"odinger representation.

\subsection{Expression of the constraint operator}
\label{s2.3}

We can now collect the results of the last two sub-sections to
obtain the quantum constraint operator starting from
the classical expression (\ref{hc1}). We have:
\ba\label{A} \hat C_{\mathrm{grav}}&=& \nonumber  \, \sin(\mb c)
\, \,\bigg[ \f{24 i\, \mathrm{sgn}(\mu)}{8 \pi \gamma^3 \mb^3
\lp^2} \,\left(\sinm \hat V \cosm - \cosm \hat V \sinm
\right)\bigg] \sin(\mb c)\\
&=:& \sin(\mb c) \,\hat A \,\sin(\mb c) \ea
where, for clarity of visualization, we have suppressed hats over
the operators $\sin(\mb c/2), \cos(\mb c/2)$ and
$\mathrm{sgn}(\mu)/\mb^3$. Let us focus on the operator $\hat{A}$
first. Some care is needed in its evaluation because, while all
other operators in this expression are densely defined,
$\mathrm{sgn}(\mu)$ is unambiguously defined only on those states
$\Phi(v)$ whose support excludes the point $v=0$. However, a
straightforward calculation shows that
\ba \f{24 i}{8 \pi \gamma^3 \mb^3 \lp^2} &\, &\left(\sinm \hat V
\cosm - \cosm \hat V \sinm \right)\, \Psi(v)\,\nonumber\\
&=\,& - \f{27K}{4} \,\sqrt{\f{8\pi}{6}} \, \f{\lp}{\gamma^{3/2}}
\,|v| \,\,\, \big[ |v - 1| - |v + 1| \big]\, \,\, \Psi(v) ~. \ea
(The expression in round brackets (containing the trignometric
functions and $\hat{V}$) is already diagonal in the $\ket{v}$
basis. Hence there is no factor ordering problem between this part
of the operator and the pre-factor involving $1/|\mb|^3$.) Since
the right hand side vanishes at $v=0$, it is in the domain of
$\mathrm{sgn}(\mu)$, whence $\hat{A}$ is well defined and given by
\be \hat A \,\Psi(v)\, =\, - \f{27K}{4} \,\sqrt{\f{8\pi}{6}} \,
\f{\lp}{\gamma^{3/2}} \,|v| \,\,\, \big| |v - 1| - |v + 1| \big|\,
\,\, \Psi(v) ~. \ee
Thus, $\ket{v}$ is an eigenket of $\hat{A}$. Since the eigenvalues
of $\hat{A}$ are real and negative, it is a negative definite
self-adjoint operator on $\Hkg$. The form of $\hat{C}_{\rm grav}$
in (\ref{A}) immediately implies that it is also self-adjoint and
negative definite on $\Hkg$. Its action is given by:
\be \label{hc2}\hat C_{\mathrm{grav}} \, \Psi(v) = f_+(v) \,\Psi(v
+ 4) + f_o(v) \, \Psi(v) + f_-(v) \, \Psi(v - 4) \ee
with
\ba f_+(v) &=& \nonumber \f{27}{16} \,\sqrt{\f{8\pi}{6}} \,
\f{K\lp}{\gamma^{3/2}} \, |v + 2|\, \,\, \big| |v + 1|
- |v + 3| \big| \\
f_-(v) &=& \nonumber f_+(v - 4)\\
f_o(v) &=&  - f_+(v) - f_-(v)~.
\ea
Thus, the new gravitational constraint is again a difference
operator. However, whereas the operator used so far in the
literature involves steps which are constant ($4\mu_o$) in
magnitude in the eigenvalues of $\hat{p}$, the new constraint
involves steps which are constant in eigenvalues of the
\emph{volume} operator $\hat{V}$. In the $\ket{\mu}$ basis these
steps vary, becoming smaller for large $\mu$. Note also that
although $\mb$ diverges at $v=0$, individual operators entering
the constraint ---and hence the full constraint operator itself---
are well-defined on the state $\ket{v=0}$.

Finally, to write the complete constraint operator we also need
the matter part of the constraint. For the massless scalar field,
in the classical theory it is given by:
\be C_{\rm matt} = 8\pi G\,\, |p|^{-\f{3}{2}}\, p_\phi^2 \ee
Thus, as usual, the non-trivial part in the passage to quantum
theory is the function $|p|^{-3/2}$. However, as with the co-triad
operator (\ref{cotriad}), this can be used by the method
introduced by Thiemann in the full theory \cite{tt,ttbook}. In
this quantization, there are two ambiguities \cite{mb-amb,mbrev},
labelled by an half integer $j$ and a real number $\ell$ in the
range $0<\ell<1$. As in \cite{aps2}, following the general
considerations in \cite{kv,ap}, we will set $j =1/2$. For $\ell$,
a general selection criterion is not available and values
$\ell=1/2$ and $\ell =3/4$ have been used varyingly in the
literature. Qualitative features of our results do not depend on
this choice. Since $\ell=1/2$ makes expressions simpler and since
we used $\ell= 3/4$ in \cite{aps2}, for simplicity and variety we
will set $\ell =1/2$ in this paper. Then, the point of departure
is the identity
\be |p|^{-\f{1}{2}} = {\mathrm{sgn}(p)}\, \bigg[\f{4}{8 \pi \lp^2
\gamma \bar \mu } {\Tr}\, \sum_k \, \tau^{k} \,\hkm \{\hkmi,\,
V^{1/3}\} \bigg] \ee
on the classical phase space. In the LQC literature, this identity
is generally used with a constant $\mu_o$ in place of $\mb$.
However, since volume depends only on $p$, in the Poisson bracket
only the derivative with respect to $c$ of the holonomy appears.
Therefore, the identity continues to hold although $\mb$ is a
function of $\mu$.
\footnote{Indeed, one could replace it by any other suitably
regular function of $\mu$. We will refrain from doing so because
the function $\mb$ was already determined for us in the
gravitational part of the Hamiltonian and choosing another
function in an ad-hoc manner for the matter part would make the
operator less natural. }

Again in the passage to quantum theory, some care is needed
because of the presence of $\mathrm{sgn}(p)$ on the right side.
However, again the image of the operator defined by the expression
in the square brackets is in the domain of $\mathrm{sgn}(p)$,
whence the total operator on the right side is densely defined:
\be \widehat{|p|^{-\f{1}{2}}}\, \Psi(v) = \f{3}{2}\left(\f{6}{8
\pi \gamma \lp^2}\right)^{1/2} \, K^{1/3} \, |v|^{1/3} \, \big|\,
|v + 1|^{1/3} - |v - 1|^{1/3} \,\big|\,\, \Psi(v) \ee
For $|v|\gg 1$, the eigenvalue is given by $(6/8\pi \g
\lp^2)^{1/2}\, {\rm sgn}(\mu)\, |\mu|^{-1/2}\,(1 + O(1/|\mu|^4))
\sim {\rm sgn}(p)\,|p|^{-1/2}$, whence the classical behavior is
recovered. However, the operator is well-behaved on the ket
$\ket{v=0}$; in fact, as is usual in LQC it is an eigenvector and the eigenvalue vanishes.

Since $\widehat{1/{\sqrt{|p|}}}$ is a well-defined, self-adjoint
operator, we can take its cube to obtain the action of
$\widehat{1/|p|^{3/2}}$. It is diagonal in the $v$ representation,
with action:
\be \label{inversevol}  \widehat{{|p|^{-\f{3}{2}}}} \Psi(v) =
\left(\f{6}{8 \pi \gamma \lp^2}\right)^{3/2}\, B(v)\, \Psi(v) \ee
where
\be \label{B} B(v) = \left(\f{3}{2}\right)^3 \, K\,\, |v| \,
\bigg| |v + 1|^{1/3} - |v - 1|^{1/3} \bigg|^3 \ee

Collecting these results we can express the total constraint
\be \hat{C}\, \Psi(v) = \left(\hat{C}_{\rm grav} + \hat{C}_{\rm
matt}\right)\, \Psi(v) = 0\, ,\ee
as follows:
\ba \label{hc3} \p^2_\phi \Psi(v,\phi)  &=& \nonumber  [B(v)]^{-1}
\, \left(C^+(v)\, \Psi(v+4,\phi) + C^o(v) \, \Psi(v,\phi)
+C^-(v)\, \Psi(v-4,\phi)\right)\\
&=:& - \Theta \,\Psi(v,\phi) ~ \ea
where the coefficients $C^\pm(v)$ and $C^o(V)$ are given by:
\ba \label{C} C^+(v) &=& \nonumber \f{3\pi K G}{8} \, |v + 2|
\,\,\,
\big| |v + 1| - |v +3|  \big|  \\
C^-(v) &=& \nonumber C^+(v - 4) \\
C^o(v) &=& - C^+(v) - C^-(v) ~. \ea
This is the Hamiltonian constraint we will work with in the
remainder of the paper. As discussed in \cite{aps2}, the form of
this constraint is similar to that of a massless Klein-Gordon
field in a static space-time, with $\phi$ playing the role of time
and the difference operator $\Theta$ of the spatial Laplace-type
operator. Hence, the scalar field $\phi$ can again be used as
`emergent time' in the quantum theory. We will examine the
operator $\Theta$ in some detail in sections \ref{s4} and
\ref{s5}. Finally, in the above construction we made a factor
ordering choice, the most significant of which is to write the
gravitational part of the constraint as $\sin\mb c \hat{A} \sin\mb
c$ by splitting the $\sin^2 \mb c$ term \cite{wk} (see (\ref{hc1})
and (\ref{A})). Since $\hat{A}$ is self-adjoint and negative
definite (and $\sin \mb c$ is self-adjoint), this ordering
directly endows the same properties on the gravitational
constraint $\hat{C}_{\rm grav}$. However, there are other
possibilities which may well  be better suited in more
complicated models. For the model under consideration, we
considered one other natural candidate in \cite{aps2} and found
that the main results were not affected by the change. Therefore
we did not use other factor orderings in the numerical simulations
with the new constraint.


\section{The Wheeler-DeWitt theory}
\label{s3}

In this section we will discuss the \WDW limit of LQC in which
effects specific to quantum geometry in the difference equation
(\ref{hc3}) are ignored. Although this limiting theory is
straightforward, we present it in some detail because it provides
a simpler and more familiar setting for constructing the physical
Hilbert space, Dirac observables and semi-classical states and
because this discussion will enable us to compare and contrast the
\WDW theory with LQC in detail.

The section is divided into two parts. In the first we obtain the
\WDW limit of (\ref{hc3}) and its general solution. In the second
we construct the physical sector of the theory, using the scalar
field $\phi$ as emergent time and show that the big bang
singularity is not resolved in the \WDW limit.

\subsection{The \WDW constraint and its general solution}
\label{s3.1}

To obtain the \WDW limit of (\ref{hc3}) we will follow the same
procedure that we used in \cite{aps2}. In particular, since the
geometrodynamical description is insensitive to the choice of
triad orientation ---i.e., to the sign of $v$--- we will restrict
ourselves to wave functions $\Psi(v)$ which are symmetric under $v
\rightarrow -v$. As in \cite{aps2}, the \WDW analogs of the LQC
quantities will be written with an underbar.

Let us begin by setting:
\be \label{chi} \chi(v) := \left(|v|\,\, \big| |v-1|-|v+1| \big|
\,\right)\,\, \left(\Psi(v+2) - \Psi(v-2)\right)\, , \ee
so that the gravitational part (\ref{hc2}) of the Hamiltonian
constraint can be written as
\be  \h{C}_{\rm grav}^{\mb}\, \Psi(v) = \f{27K\lp}{16}\,
\sqrt{\f{8\pi}{6\g^3}}\,\, [\chi(v+2) -\chi(v-2)]\, .\ee
To obtain the \WDW limit, let us assume $|v| \gg 1$ and $\Psi(v)$
is smooth.  Then, we have:
\be \h{C}_{\rm grav}^{\mb}\, \Psi(v) = 54 K\lp\,
\sqrt{\f{8\pi}{6\g^3}}\,\, \f{\dd}{\dd v} (|v| \f{\dd \Psi(v)}{\dd
v}) + O(v^{n-3}\f{\dd^n\Psi}{\dd v^n}) \ee
where $n\ge 3$. Thus, if we restrict ourselves to wave functions
$\Psi$ which are slowly varying in the sense that the second term
on the right hand side is negligible compared to the first, we
obtain the \WDW limit of the gravitational part of the constraint:
\be \label{wdw1} \h{C}_{\rm grav}^{\rm wdw} \Psi(v) = 54 K\lp\,
\sqrt{\f{8\pi}{6\g^3}}\,\, \f{\dd}{\dd v} \left(|v| \f{\dd
\Psi(v)}{\dd v}\right).\ee
This approximation is not uniform because the terms which are
neglected depend on $\Psi$. We will show at the end that these
assumptions are realized in a self-consistent manner on
semi-classical states of interest. For now we only note that
$\h{C}_{\rm grav}^{\rm wdw}$ is self-adjoint and negative definite
on the kinematic Hilbert space $\Hkwdw = L^2(\R, dv)$ of the \WDW
theory.

Finally, recall that the full constraint is given by
\be \hat{C} = \h{C}_{\rm grav} + 8\pi G\, \left(\f{6}{8\pi\g
\lp^2}\right)^{3/2}\, B(v) \h{p}_\phi^2 \ee
where the function $B(v)$ is defined in (\ref{B}). This function
is well-approximated by $\ub{B}(v):= K |v|^{-1}$ for $|v| \gg 1$.
Hence, the \WDW limit of the full constraint is given by:
\ba \label{wdw2} \p^2_\phi\,\, \ul\Psi(v,\phi) &=& {12\pi G}\,\,
v\p_v
\big(v\p_v\, \ul\Psi(v,\phi)\big) \nonumber\\
&=:& -\ul\Theta\,\,  \ul\Psi(v,\phi) \ea
The operator $\ul\Theta$ commutes with the `parity' operator $\Pi$
which flips the triad orientation: $\Pi\, \ul\Psi(v) =
\ul\Psi(-v)$.%
\footnote{In the full classical theory, the orientation reversal
is a phase space symmetry because both triad $e^a_i$ and its
conjugate variable $k_a^i := k_{ab} e^{bi}$ change signs under
this transformation. In the present model, this corresponds to the
symplectomorphism $(c,p) \rightarrow (-c,-p)$ which is then
naturally lifted to an automorphism of the  quantum algebra. The
mapping $\Pi$ is the induced action of this automorphism on
states.}
Hence it preserves the space of wave functions under
consideration, namely the eigenspace of $\Pi$ with eigenvalue
$+1$.

Note that if, as is in geometrodynamics, we were to represent
quantum states as functions $\t{\ul\Psi}(a,\phi)$, the operator
$\ul\Theta$ would have the action: $\ul\Theta\, \t{\ul\Psi} =
(4\pi G/3)\, (a\p /\p a) (a\p/\p a)(\t{\ul\Psi})$. Now, in
geometrodynamics, the classical constraint has the form
$G^{AB}p_Ap_B =0$ where $G^{AB}$ is the DeWitt metric on the
2-dimensional configuration space spanned by the pair $(a,\phi)$.
In quantum theory, the natural choice of factor ordering is to use
the Laplace-Beltrami operator of $G^{AB}$ \cite{ag}. Then, the
\WDW equation has the form: $\p_\phi^2 \t{\ul\Psi} = (4\pi G/3)\,
(a\p/\p a) (a\p/\p a)\t{\ul\Psi}$ \cite{ck}, which is precisely
our equation (\ref{wdw2}). Thus the \WDW limit of the current,
`improved' Hamiltonian constraint, automatically yields the
`natural' factor ordering of the \WDW theory. This provides an
indirect support for the factor ordering in LQC that led us to
(\ref{hc3}).

Since the form of the \WDW limit is symmetric in $\phi$ and $v$ we
could use either the volume of the universe or the scalar field as
the `emergent time' and the other variable as the true dynamical
degree of freedom. However, as noted in \cite{aps1,aps2}, in the
(k=1) closed model, $\phi$ is better suited as the time variable.
More importantly,  the form of the LQC constraint (\ref{hc3}) is
such that it would be technically difficult to regard $v$ as time
beyond the \WDW limit. Therefore, we will regard $\phi$ as time so
that the dynamics is described by the evolution of the volume of
the universe with respect to this emergent time, $\phi$. As
emphasized in \cite{aps2}, it is \emph{not} essential to make a
specific choice of an `emergent time'; one can construct the
physical sector of the theory without making a choice. However,
choosing $\phi$ as time provides a heuristic understanding of the
intermediate stages of the procedure and lets us interpret the
final results in the language of the more familiar notion of
`evolution' rather than in terms of a `frozen formalism'.

The form of the \WDW equation (\ref{wdw2}) is very simple; if we
set $x\!=\!\ln v$, the \WDW constraint assumes the form of a
massless Klein Gordon equation in the flat space spanned by $x$
and $\phi$. However, to bring out similarities and contrasts
between this \WDW theory and the LQC description presented in sections
\ref{s4} and \ref{s5}, we will use the same steps as the ones used
there.

Let us first note that the operator $\ul\Theta$ is positive
definite and self-adjoint on the Hilbert space $L^2_s(\R,
K^{-1}\ul{B}(v)\dd v)$ where the subscript $s$ denotes the
restriction to the symmetric eigenspace of $\Pi$. Its
eigenfunctions $\ub{e}_k$ with eigenvalue $\omega^2 (\ge 0)$ are
2-fold degenerate on this Hilbert space. Therefore, they can be
labelled by a real number $k$:
\be \label{ek_eq} \ub{e}_k(v) := \f{1}{\sqrt{2\pi}} \, e^{ik\ln |v|}\ee
where $k$ is related to $\omega$ via $\omega= \sqrt{12\pi G}|k|$.
On $L^2_s(\R, K^{-1}\ul{B}(v)\dd v)$, they form an orthonormal
basis: $(e_k, e_k') = \delta(k,k')$. A general solution to
(\ref{wdw2}) with initial data in the Schwartz space of rapidly
decreasing functions can be written as
\be \label{sol1} \ul\Psi(v, \phi) = \int_{-\infty}^{\infty}\, \dd
k \, \t\Psi_{+}(k)\, \ub{e}_k(v) e^{i\omega \phi}+ \t\Psi_{-}(k)\,
\bar{\ub{e}}_k(v) e^{-i\omega \phi}\,  \ee
for some $\t\Psi_\pm(k)$ also in the Schwartz space. Following the
terminology generally used in the Klein-Gordon theory, the
solution will be said to be `incoming' (`contracting') if
$\t\Psi_\pm(k)$ have support only on the positive half of the $k$
axis and `outgoing' or (`expanding') if they have support only on
the negative half of the $k$ axis. If $\t\Psi_-(k)$ vanishes, the
solution will be said to be of positive frequency and if
$\t\Psi_+(k)$ vanishes it will be said to be of negative
frequency.

As usual, the positive and negative frequency solutions satisfy
first order `evolution' equations, obtained by taking a
square-root of the constraint (\ref{wdw2}):
\be \label{freq1}\mp i\, \p_\phi \ul\Psi(v,\phi) =
\sqrt{\ul\Theta}\, \ul\Psi(v,\phi)\, . \ee
If $f(v)$ is the initial data for these equations at `time
$\phi=\phi_o$', the solutions are given by:
\be \label{sol2} \ul\Psi_\pm(v,\phi) = e^{\pm i \sqrt{\ul\Theta}\,
\,(\phi-\phi_o)}\,\, f(v) \ee
\subsection{Physics of the \WDW theory}
\label{s3.2}

Solutions (\ref{sol1}) to the \WDW equation are not normalizable
in $\Hkwdw$ (because zero is in the continuous part of the
spectrum of the \WDW operator). Our first task is to endow the
space of these physical states with a Hilbert space structure.
There are several possible avenues. As in \cite{aps2}, we will
begin with one that is somewhat heuristic but has direct physical
motivation. The idea \cite{aabook,at} is to introduce operators
corresponding to a complete set of Dirac observables and select
the required inner product by demanding that they be self-adjoint.
In the classical theory, since $p_\phi$ is a constant of motion,
it is a Dirac observable. While $v$ is not a constant of motion,
on each dynamical trajectory $v(\phi)$ is a monotonic function of
$\phi$, whence $v\mid_{\phi=\phi_o}$ is a Dirac observable for any
fixed $\phi_o$. These form a complete set and it is
straightforward to write the corresponding quantum operators.
However, because we are interested only in states $\Psi(v,\phi)$
which are symmetric under $v \rightarrow -v$, it suffices to
consider $|v|_{\phi_o}$ in place of $v|_{\phi_o}$. Now, since
$\hat{p}_\phi$ commutes with the \WDW operator in (\ref{wdw2}),
given a (symmetric) solution $\ul\Psi(v,\phi)$ to (\ref{wdw2}),
\be \label{dirac1}\hat{p}_\phi\, \ul\Psi(v, \phi) := -i\hbar
\f{\partial \ul\Psi}{\partial \phi} \ee
is again a (symmetric) solution. So, we can just retain this
definition of $\hat{p}_\phi$ from $\Hkwdw$. The Schr\"odinger type
evolutions (\ref{sol2}) enable us to define the other Dirac
observable $|\hat{v}|_{\phi_o}$: Given a (symmetric) solution
$\ul\Psi(v,\phi)$ to (\ref{wdw2}), we can first decompose it into
positive and negative frequency parts $\ul\Psi_\pm(v,\phi)$,
freeze them at $\phi=\phi_o$, multiply this `initial datum' by
$|v|$ and evolve via (\ref{sol2}):
\be \label{dirac2} {|\h{v}|_{\phi_o}}\,\ul\Psi (v,\,\phi) =
e^{i\sqrt{\ul{\Theta}}\,\,(\phi-\phi_o)}\, |v| \,
\ul\Psi_+(v,\,\phi_o) +
e^{-i\sqrt{\ul{\Theta}}\,\,(\phi-\phi_o)}\, |v| \,
\ul\Psi_-(v,\,\phi_o) \ee
The result is again a (symmetric) solution to the \WDW equation
(\ref{wdw2}). Furthermore, both these operators preserve the
positive and negative frequency subspaces. Since they constitute a
complete family of Dirac observables, we have
\emph{superselection}. In quantum theory we can restrict ourselves
to one superselected sector. In what follows, for definiteness
\emph{we will focus on the positive frequency sector and, from now
on, drop the suffix $+$}.

We now seek an inner product on the space of positive frequency
solutions $\ul\Psi(v,\phi)$ to (\ref{sol2}) (invariant under the
$v$ reflection) which makes $\hat{p}_\phi$ and
$|\hat{v}|_{\phi_o}$ self-adjoint. Each of these solutions is
completely determined by its initial datum $\ul\Psi(v,\phi_o)$ and
the Dirac observables have the following action on the datum:
\be {|\h{v}|_{\phi_o}}\, \ul\Psi (v,\phi) = |v|\,
\ul\Psi(v,\phi_o),\quad {\rm and} \quad \hat{p}_\phi\,
\ul\Psi(v,\phi_o) = \hbar \sqrt{\ul{\Theta}}\,\,
\ul{\Psi}(v,\phi_o)\, . \ee
Therefore, it follows that (modulo an overall rescaling,) the
unique inner product which will make these operators self-adjoint
is just:
\be \label{ip1} \langle\ul\Psi_1\,|\ul\Psi_2\, \rangle_{\rm phy} =
\int_{\phi=\phi_o}\!\! dv\, \ub{B}(v)\, \bar{\ul\Psi}_1(v,\phi)\,
\ul\Psi_2(v,\phi) \ee
(see e.g. \cite{aabook,at}). Note that the inner product is
conserved, i.e., is independent of the choice of the `instant'
$\phi=\phi_o$. Thus, \emph{the physical Hilbert space $\Hpwdw$ is
the space of positive frequency wave functions $\ul\Psi(v,\phi)$
which are symmetric under the $v$ reflection and have a finite
norm, defined by (\ref{ip1}).} The procedure has already provided
us with a representation of our complete set of Dirac observables
on this $\Hpwdw$:
\be \label{dirac3}{|\h{v}|_{\phi_o}}\,\, \ul\Psi(v,\phi) =
e^{i\sqrt{\ul{\Theta}}\,\,(\phi-\phi_o)}\, |v|
\,\ul\Psi(v,\phi_o),\quad {\rm and} \quad \hat{p}_\phi\,
\ul\Psi(v,\phi) = \hbar \sqrt{\ul{\Theta}}\, \ul\Psi(v,\phi)\, .
\ee
Arguments of \cite{aps2} can be repeated to show that the same
representation of the algebra of Dirac observables can be obtained
by the group averaging method \cite{dm,hm2} which is
mathematically more stream-lined.

Finally, using the physical Hilbert space and this complete set of
Dirac observables we can now introduce semi-classical states and
study their evolution. Let us fix an `instant of time'
$\phi=\phi_o$ and construct a semi-classical state which is peaked
at $p_\phi = p_\phi^\star$ and $v|_{\phi_o} = v^\star$. Since we
would like the peak to be at a point that represents a large
classical universe, we are led to choose $v^\star \gg 1$ and (in
the natural classical units $c$=$G$=1)\,\, $p_\phi^\star \gg
\hbar$. In the closed ($k\!=\! 1$) models for example, the second
condition is necessary to ensure that the universe expands out to
a size much larger than the Planck scale. At `time' $\phi=\phi_o$,
consider the state
\be \label{sc} \ul{\Psi}(v,\phi_o) = \int_{-\infty}^\infty dk\,\,
\t{\Psi}(k)\,\, \ub{e}_k(v)\, e^{i\omega\,(\phi_o-\phi^\star)},
\quad {\rm where}\,\, \t{\Psi}(k) =
e^{-\f{(k-k^\star)^2}{2\sigma^2}}\, . \ee
Here $k^\star = - p_\phi^\star/\sqrt{12\pi G\hbar^2}$ and
$\phi^\star = - \ln |v^\star|/\sqrt{1/12\pi G}\, + \phi_o$. It is
easy to evaluate the integral in the approximation $\omega =
-\sqrt{12\pi G}\, k$ ---which is justified because $\t\Psi(k)$ is
sharply peaked at $k^\star$ and $k^\star \ll -1$--- and calculate
mean values of the Dirac observables and their fluctuations. One
finds that, as required, at $\phi =\phi_o$ the state is sharply
peaked at values $v^\star, p_\phi^\star$. The above construction
is closely related to that of coherent states in non-relativistic
quantum mechanics. The main difference is that the observables of
interest are not $v$ and its conjugate momentum but rather $v$ and
$p_\phi$ ---the momentum conjugate to `time', i.e., the analog of
the Hamiltonian in non-relativistic quantum mechanics.

We can now ask for the evolution of this state. Does it remain
peaked at the classical trajectory defined by $p_\phi =
p_\phi^\star$ and passing through $v=v^\star$ at $\phi =\phi_o$?
This question is easy to answer because (\ref{sol2}) implies that
the (positive frequency) solution to $\ul{\Psi}(v,\phi)$
(\ref{wdw2}) defined by the initial data (\ref{sc}) is obtained
simply by replacing $\phi_o$ by $\phi$ in (\ref{sc})! Since
$\sigma$, the measure of dispersion in (\ref{sc}), does not depend
on $\phi$, it follows that $\ul{\Psi}(v,\phi)$ continues to be
peaked at a trajectory
\be \phi =  \sqrt{\f{1}{12\pi G}}\,\,  \ln{\f{|v|}{|v^\star|}} +
\phi_o\, \ee
which is precisely the classical solution of interest. This is
precisely what one would hope during the epoch in which the
universe is large. However, the property holds also in the Planck
regime and, in the backward evolution, the semi-classical state
simply follows the classical trajectory into the big-bang
singularity. (Had we worked with positive $k^\star$, we would have
obtained a contracting solution and then the forward evolution
would have followed the classical trajectory into the big-crunch
singularity.) In this sense, the \WDW evolution does not resolve
the classical singularity.

We will show in sections \ref{s4} and \ref{s5} that the situation
is very different in LQC. This can occur because the \WDW equation
is a good approximation to the discrete equation only for large
$v$. Furthermore, as noted before, the approximation is not
uniform but depends on the state: in arriving at the \WDW equation
from LQC we had to neglect $\Psi$ dependent terms of the form
$O(v^{n-3}\f{\dd^n\Psi}{\dd v^n})$ for $n\ge 3$. For
semi-classical states considered above, this implies that the
approximation is excellent for $v \gg k^\star$ but becomes
inadequate when the peak of the wave function lies at a value of
$v$ comparable to $k^\star$. Then, the LQC evolution departs
sharply from the \WDW evolution. We will find that, rather than
following the classical trajectory into the big bang singularity,
the peak now exhibits a bounce. Since large values of $k^\star$
are classically preferred, the value of $v$ at the bounce can be
quite large. However, as remarked in Sec. \ref{s1}, we will find
that the matter density at the bounce point is comparable to the
Planck density, independent of the value of $k^\star$.

\textbf{Remark:} In the above discussion for simplicity we
restricted ourselves to eigenfunctions $\ub{e}_k(v)$ which are
symmetric under $v\,\rightarrow\, -v$ from the beginning. Had we
dropped this requirement, we would have found that there is a
4-fold (rather than 2-fold) degeneracy in the eigenfunctions of
$\ul{\Theta}$. Indeed, if $\theta(v)$ is the step function
($\theta(v) =0$ if $v < 0$ and $\, =1$ if $v >0$), then
$\theta(v)\ub{e}_{|k|},\, \theta(v) \ub{e}_{-|k|},\,
\theta(-v)\ub{e}_{|k|},\, \theta(-v) \ub{e}_{-|k|}$ are all
continuous functions of $v$ which satisfy the eigenvalue equation
(in the distributional sense) with eigenvalue $\omega^2 = 12\pi G
\,k^2 $. This fact will be relevant in the next section.

\section{Analytical issues in Loop quantum cosmology}
\label{s4}

We will now analyze the model using LQC. Since the form of the LQC
`evolution equation' is very similar to that of the \WDW theory,
we will be able to construct the physical Hilbert space and Dirac
observables following the ideas introduced in section \ref{s3.2}.

\subsection{Emergent time and the general solution to the LQC
Hamiltonian constraint} \label{s4.1}

Recall that the LQG Hamiltonian constraint is given by Eq
(\ref{hc3}):
\ba \label{qh7} \p^2_\phi \Psi(v,\phi)  &=& \nonumber  [B(v)]^{-1} \,
\left(C^+(v)\, \Psi(v+4,\phi) + C^o(v) \, \Psi(v,\phi)
+C^-(v)\, \Psi(v-4,\phi)\right)\\
&=:& - \Theta \,\Psi(v,\phi) ~ \ea
where the coefficients $C^\pm, C^o$ are given by (\ref{C}).%
\footnote{Note that this fundamental evolution equation makes no
reference to the Barbero-Immirzi parameter $\gamma$ or the area
gap $\Delta$. This is the equation used in numerical simulations.
To interpret the results in terms of scale factor, however, values
of $\gamma$ and $\Delta$ become relevant.}
Since the operator $\Theta$ acts only on the argument $v$ of
$\Psi(v,\phi)$, we have a neat separation of variables. The form
of the constraint now suggests that it is natural to regard $\phi$
as emergent time. To implement this idea, let us introduce an
appropriate kinematical Hilbert space for both geometry and the
scalar field: $\Hk := L^2(\R_{\rm Bohr}, B(v)\dd\mu_{\rm Bohr})
\otimes L^2(\R, \dd \phi)$. Since $\phi$ is to be thought of as
`time' and $v$ as the genuine, physical degree of freedom which
evolves with respect to this `time', we chose the standard
Schr\"odinger representation for $\phi$ but the `polymer
representation' for $v$ to correctly incorporate the quantum
geometry effects. This is a conservative approach in that the
results will directly reveal the manifestations of quantum
geometry. Had we chosen a non-standard representation for the
scalar field, these effects would have been mixed with those
arising from an unusual representation of `time evolution' and,
furthermore, comparison with the \WDW theory would have become
more complicated. (However, the use of a `polymer representation'
for $\phi$ may become necessary to treat inhomogeneities in an
adequate fashion.)

The form of (\ref{hc3}) is the same as that of the \WDW constraint
(\ref{wdw2}) and $\Theta$ is again a positive self-adjoint
operator, now on $L^2(\R_{\rm Bohr}, B(v)\dd\mu_{\rm Bohr})$. The
main difference is that while the \WDW $\ul\Theta$ is a
differential operator, the LQC $\Theta$ is a \emph{difference}
operator. This gives rise to certain technically important
distinctions. For, now the space of physical states ---i.e. of
appropriate solutions to the constraint equation--- is naturally
divided into sectors each of which is preserved by the `evolution'
and by the action of our Dirac observables. Thus, there is
super-selection. Let $\La_{|\epsilon|}$ denote the `lattice' of
points $\{|\epsilon|+4n,\, n\in \Z\}$ on the $v$-axis,
$\La_{-|\epsilon|}$ the `lattice' of points $\{-|\epsilon|+4n,\,
n\in \Z\}$ and let $\La_{\epsilon} = \La_{|\epsilon|} \cup
\La_{-|\epsilon|}$ where as usual $\Z$ denotes the set of integers. Let $\H_{|\epsilon|}^{\rm
grav},\H_{-|\epsilon|}^{\rm grav}$ and $\H_{\epsilon}^{\rm grav}$
denote the subspaces of $L^2(\R_{\rm Bohr}, B(v) \dd\mu_{\rm
Bohr})$ with states whose support is restricted to lattices
$\La_{|\epsilon|}, \La_{-|\epsilon|}$ and $\La_\epsilon$. Each of
these three subspaces is mapped to itself by $\Theta$ which is 
self-adjoint and positive definite on all three Hilbert spaces.

However, since $\H_{|\epsilon|}^{\rm grav}$ and
$\H_{-|\epsilon|}^{\rm grav}$ are mapped to each other by the
parity operator $\Pi$, only $\H_\epsilon^{\rm grav}$ is left
invariant by $\Pi$. Now, because $\Pi$ reverses the triad
orientation, it represents a large gauge transformation. In gauge
theories, we have to restrict ourselves to sectors, each
consisting of an eigenspace of the group of large gauge
transformations. (In QCD in particular this leads to the $\theta$
sectors.) The group generated by $\Pi$ is just $\Z_2$, whence
there are only two eigenspaces, with eigenvalues $\pm 1$. Since
there are no fermions in our theory, there are no parity violating
processes whence we are led to choose the symmetric sector with
eigenvalue $+1$. 
Thus, we are primarily interested in the symmetric subspace of
$\H_\epsilon^{\rm grav}$; the other two Hilbert spaces will be
useful only in the intermediate stages of our discussion.

Our first task is to explore properties of the operator $\Theta$.
Since it is self-adjoint and positive definite, its spectrum is
non-negative. Therefore, as in the \WDW theory, we will denote its
eigenvalues by $\omega^2$. Let us first consider a generic
$\epsilon$, i.e., not equal to $0$ or $2$, so that the lattices
$\La_{\pm|\epsilon|}$ are distinct. On each of the two Hilbert
spaces $\H_{\pm|\epsilon|}^{\rm grav}$, we can solve for the
eigenvalue equation $\Theta\, e_\omega(v) = \omega^2\, e_\omega
(v)$, i.e.,
\be  \label{eq:eigen} C^+(v)e_\omega(v+4) + C^o(v)e_\omega(v) +
C^-(v) e_\omega(v-4)  = \omega^2 B(v) e_\omega(v)\, \ee
Since this equation has the form of a recursion relation and since
the coefficients $C^\pm(v)$ never vanish on the `lattices' under
consideration, it follows that we will obtain an eigenfunction by
freely specifying, say,  $\Psi(v^\star)$ and $\Psi(v^\star+4)$ for
any $v^\star$ on the lattice ${\cal L}_{|\epsilon|}$ or ${\cal
L}_{-|\epsilon|}$. Hence the eigenfunctions are 2-fold degenerate
on each of $\H_{|\epsilon|}^{\rm grav}$ and $\H_{-|\epsilon|}^{\rm
grav}$. On $\H_\epsilon^{\rm grav}$, therefore, the eigenfunctions
are 4-fold degenerate as in the \WDW theory. Consequently
$\H_\epsilon^{\rm grav}$ admits an orthonormal basis $e_\omega^I$
where the degeneracy index $I$ ranges from $1$ to $4$, such that
\be \langle e_\omega^I|e_{\omega'}^{I'}\rangle\,\, =\,\,
\delta_{I,I'}\, \delta(\omega,\, \omega'). \ee
(The Hilbert space $\H_\epsilon^{\rm grav}$ is separable and the
spectrum is equipped with the standard topology of the real line.
Therefore we have the Dirac distribution
$\delta(\omega,\,\omega')$ rather than the Kronecker delta
$\delta_{\omega,\,\omega'}$.) As usual, every element $\Psi(v)$ of
$\H_\epsilon^{\rm grav}$ can be expanded as:
\be \Psi (v) = \int_0^\infty d\omega\, \t\Psi_I(\omega)
e^I_{\omega}(v) \quad {\rm where}\,\,\,\, \t\Psi_I(\omega)\, =\,
\langle e^I_\omega|\Psi\rangle\, , \ee
The numerical analysis of section \ref{s5} and comparison with the
\WDW theory are facilitated by making a convenient choice of this
basis in $\H_\epsilon^{\rm grav}$, i.e., by picking specific
vectors from each 4 dimensional eigenspace spanned by
$e^I_\omega(v)$. To do so, note first that, as one might expect,
every eigenvector $e_\omega^I(v)$ has the property that it
approaches specific eigenvectors of the \WDW differential operator
$\ul\Theta$ as $v \rightarrow \pm \infty$. (The precise rate of
approach is discussed in section \ref{s5.1}.) The idea is to use
this fact to synchronize the basis in LQC with that in the \WDW
theory. However, in general the two limiting \WDW eigenfunctions
are distinct. Indeed, because of the nature of the \WDW operator
$\ul\Theta$, its eigenvectors can be chosen to vanish on the
entire negative (or positive) $v$-axis; their behavior on the two
half lines is uncorrelated. (See the remark at the end of section
\ref{s3.2}.) Eigenvectors of the LQC operator $\Theta$ on the
other hand are rigid; their values at any two lattice points
determine their values on the entire lattice
$\La_{\pm|\epsilon|}$. Therefore, the synchronization can be done
only at one end and we choose to do so at $v=\infty$. As in the
\WDW theory, let us introduce a real variable $k$ satisfying
$\omega^2 = (12\pi G) k^2$ and use $k$ in place of $\omega$ to
label the orthonormal basis. Then the basis of interest will be the following:
\begin{quote}
i) Denote by $e_{-|k|}^\pm(v)$ the basis vector in
$\H_{\pm|\epsilon|}^{\rm grav}$ with eigenvalue $\omega^2$, which
is proportional to the \WDW $\ub{e}_{-|k|}(v)$ in the limit $v
\rightarrow \infty$; (i.e., it has only `outgoing' or `expanding'
component in this limit);\\
ii) Denote by  $e_{|k|}^\pm(v)$ the basis vector in
$\H_{\pm|\epsilon|}^{\rm grav}$ with eigenvalue $\omega^2$ which
is orthogonal to $e_{-|k|}^\pm(v)$. (Since eigenvectors are 2-fold
degenerate in each of $\H_{\pm|\epsilon|}^{\rm grav}$, the vector
$e_{|k|}^\pm (v)$ is uniquely determined up to a multiplicative
phase factor.)
\end{quote}
We thus have an orthonormal basis $e_k^\pm$ in $\H_{\epsilon}^{\rm
grav}$ with $k \in \R$: $\langle e_k^\pm|e_{k'}^\pm\rangle =
\delta(k,k')$, and $\langle e_k^+|e_{k'}^-\rangle = 0$. The four
eigenvectors with eigenvalue $\omega^2$ are now $e^+_{|k|},
e^+_{-|k|}$ which have support on the `lattice'
$\La_{|\epsilon|}$, and $e^-_{|k|}, e^-_{-|k|}$ which have support
on the `lattice' $\La_{-|\epsilon|}$. As we will see in section
\ref{s5.1}, this basis is well-suited for numerical analysis.

Since physical states use only the symmetric sector of
$\H_{\epsilon}$, in this analytic discussion we will be interested
only in the symmetric combinations
\be \label{eq:e-symm} e^{(s)}_k (v) = \f{1}{{2}}\, \left(e^+_k(v)
+ e^+_k(-v) + e^-_k(v) + e^-_k(-v) \right) \ee
of the basis vectors which are invariant under $\Pi$. Any
symmetric element $\Psi(v)$ of $\H_{\epsilon}^{\rm grav}$ can be
expanded as
\be \label{sym} \Psi(v) = \int_{-\infty}^{\infty} dk\,\,
\t\Psi(k)\,\, e^{(s)}_k(v) \ee

We can now write down the general symmetric solution to the
quantum constraint (\ref{hc3}) with initial data in
$\H_{\epsilon}^{\rm grav}$ :
\be \label{sol3}\Psi(v,\phi) = \int_{-\infty}^{\infty} dk\,\,
[\t\Psi_+(k) e^{(s)}_k(v) e^{i\omega\phi} + \t\Psi_- (k)
\bar{e}^{(s)}_k(v) e^{-i\omega\phi}]\ee
where $\t{\Psi}_\pm(k)$ are in $L^2(\R, dk)$. As $v  \rightarrow
\pm \infty$, these approach solutions (\ref{sol1}) to the \WDW
equation. However, the approach is not uniform in the Hilbert
space but varies from solution to solution. As mentioned in
section \ref{s3.2}, the LQC solutions to (\ref{hc3}) which are
semi-classical at late times can start departing from the \WDW
solutions for relatively large values of $v$ (more precisely, at $v \sim k$).

As in the \WDW theory, if $\Psi_-(k)$ vanishes, we will say that
the solution is of positive frequency and if $\Psi_+(k)$ vanishes
we will say it is of negative frequency. Thus, every solution to
(\ref{hc3}) admits a natural positive and negative frequency
decomposition. The positive (respectively negative) frequency
solutions satisfy a Schr\"odinger type first order differential
equation in $\phi$:
\be \label{eq:fund_eq}
\mp i\f{\p\Psi_\pm}{\p\phi} = \sqrt{\Theta} \Psi_\pm \ee
but with a Hamiltonian $\sqrt{\Theta}$ (which is non-local in
$v$). Therefore the solutions with initial datum $\Psi(v, \phi_o)
= f_\pm(v)$ are given by:
\be \Psi_\pm(v,\phi) \,=\, e^{\pm
i\sqrt{\Theta}\,\,(\phi-\phi_o)}\, f_\pm(v,\phi)\ee

So far, we considered a generic $\epsilon$. We will conclude by
summarizing the situation in the special cases, $\epsilon= 0$ and
$\epsilon =2$. In these cases, differences arise because the
individual lattices are invariant under the reflection $v
\rightarrow -v$, i.e., the lattices $\La_{|\epsilon|}$ and
$\La_{-|\epsilon|}$ coincide. As before, there is a 2-fold
degeneracy in the eigenvectors of $\Theta$ on any one lattice. For
concreteness, let us label the Hilbert spaces
$\H_{|\epsilon|}^{\rm grav}$ and choose the basis vectors
$e_k^+(v)$, with $k\in \R$ as above. Now, symmetrization can be
performed on each of these Hilbert spaces by itself. So, we have:
\be \label{sb} e_k^{(s)}(v) = \f{1}{\sqrt{2}}\, (e^+_k(v) + e^+_k
(-v))\ee
However, the vector $e_{|k|}^{(s)}(v)$ coincides with the vector
$e_{-|k|}^{(s)}(v)$ so there is only one symmetric eigenvector per
eigenvalue. This is not surprising: the original degeneracy was
2-fold (rather than 4-fold) and so there is one symmetric and one
anti-symmetric eigenvector per eigenvalue. Nonetheless, it is
worth noting that there is a precise sense in which the Hilbert
space of symmetric states is only `half as big' in these
exceptional cases as they are for a generic $\epsilon$.

For $\epsilon=2$, there is a further subtlety because $C^+$
vanishes at $v= -2$ and $C^-$ vanishes at $v= 2$. Thus, in this
case, as in the \WDW theory, there is a decoupling and the
knowledge of the eigenfunction $e^+_k(v)$ on the positive $v$-axis
does not suffice to determine it on the negative $v$ axis and
vice-versa. However, the degeneracy of the eigenvectors does not
increase but remains 2-fold because the (\ref{hc3}) now introduces
two new constraints: $C^\pm(\pm 2) e^+_k(\pm 6) = [\omega^2 B(\pm
2) - C^o(\pm 2)] e^+_k(\pm 2) =0$. Conceptually, this difference
is not significant; there is again a single symmetric
eigenfunction for each eigenvalue.

\subsection{The Physical sector}
\label{s4.2}

Results of section \ref{s4.1} show that while the LQC operator
$\Theta$ differs from the \WDW operator $\ul\Theta$ in interesting
ways, the structural form of the two Hamiltonian constraint
equations is the same. Therefore, apart from the issue of
superselection sectors, introduction of the Dirac observables and
determination of the inner product either by demanding that the
Dirac observables be self-adjoint or by carrying out group
averaging is completely analogous in the two cases. Therefore, we
will not repeat the discussion of section \ref{s3.2} but only
summarize the final structure.

The sector of the physical Hilbert space $\Hp^\epsilon$ labelled by
$\epsilon \in [0,\, 2 ]$ consists of positive frequency solutions
$\Psi(v,\phi)$ to (\ref{eq:fund_eq}) with initial data $\Psi (v, \phi_o)$
in the symmetric sector of $\H^\epsilon_{\rm grav}$. Eq.
(\ref{sol3}) implies that they have the explicit expression in
terms of our eigenvectors $e^{(s)}_k(v)$
\be \label{eq:psi-int} \Psi(v,\phi) = \int_{-\infty}^\infty dk\,
\t\Psi(k) \,e^{(s)}_k(v)\, e^{i\omega\phi}\,\, , \ee
where, as before, $\omega^2 = 12\pi Gk^2$ and $e^{(s)}_k(v)$ is
given by (\ref{eq:e-symm}) and (\ref{sb}). By choosing appropriate
functions $\t\Psi(k)$, this expression will be evaluated in
section \ref{s5.1} using Fast Fourier Transforms. The resulting
$\Psi(v,\phi)$ will provide, numerically, quantum states which are
semi-classical for large $v$. The physical inner product is given
by
\be \label{ip2} \langle\Psi_1|\Psi_2\rangle_\epsilon\,\, = \sum_{v
\in \{\pm|\epsilon| + 4n;\, n\in \Z\}}\,\, B(v)
\bar{\Psi}_1(v,\phi_o)\, \Psi_2(v, \phi_o) \ee
for any $\phi_o$. The action of the Dirac observables is
independent of $\epsilon$, and has the same form as in the \WDW
theory:
\be \label{dirac4}{|\h{v}|_{\phi_o}}\, \Psi(v,\phi) =
e^{i\sqrt{{\Theta}}\,\,(\phi-\phi_o)}\, |v| \,\Psi(v,\phi_o),\quad
{\rm and} \quad \hat{p}_\phi \Psi(v,\phi) = - \, i\hbar\, \f{\p
\Psi(v,\phi)}{\p\phi} \, . \ee
The kinematical Hilbert space $\Hk$ is non-separable but, because
of super-selection, each physical sector $\Hp^\epsilon$ is
separable. Eigenvalues of the Dirac observable
${|\h{v}|_{\phi_o}}$ constitute a discrete subset of the real line
in each sector. The set of these eigenvalues in distinct sectors
is distinct. Therefore which sector  actually occurs is a question
that can be in principle answered experimentally, provided one has
access to microscopic measurements which can distinguish between
values of the scale factor which differ by $\sim 1.3 \lp$. This
will not be feasible in the foreseeable future. Of greater
practical interest are the coarse-grained measurements, where the
coarse graining occurs at significantly greater scales. For these
measurements, different sectors would be
indistinguishable and one could work with any one.

{\bf Remark:} The procedure used in this section is quite general
in the sense that it is applicable for a large class of systems.
In this sense, the physical Hilbert spaces $\Hp^\epsilon$
constructed here are natural. However, using the special
structures available in this model, one can also construct an
inequivalent representation which is closer to that used in the
\WDW theory. The main results on the bounce also hold in that
representation. See Appendix C of \cite{aps2}.

\section{LQC: Numerical issues}
\label{s5}

In this section, we will find physical semi-classical states in
LQC and analyze their properties numerically. This section is
divided into three parts. In the first we study eigenfunctions
$e^\pm_k (\mu)$ of $\Theta$ and introduce a  method of
constructing a `general' physical state by direct evaluation of
the right side of (\ref{eq:psi-int}). In the second part we solve
the initial value problem starting from initial data at
$\phi=\phi_o$, thereby obtaining a `general' solution to the
difference equation (\ref{qh7}). In the third we summarize the
main results. Readers who are not interested in the details of
simulations can go directly to the third subsection.

A large number of simulations were performed within each of the
approaches by varying the parameters in the initial data and
working with different lattices ${\cal L}_\epsilon$ introduced in
Sec. \ref{s4}. They show the robustness of final results. To avoid
making the paper excessively long, we will only show
illustrative plots.

\subsection{Eigenfunctions of the $\Theta$ operator, direct evaluation
  of the integral representation of the state}
\label{s5.1}

Here we will establish properties of the general eigenfunctions
$e_\omega(v)$ of $\Theta$ introduced in section \ref{s4.1},
briefly present the method of explicit calculation of eigenfunctions
in the  symmetric sector and construct semiclassical states through
direct evaluation of their integral representation given by \eqref{sol3}.
This method is an exact analog of the one used in \cite{aps2} and
plays only an auxiliary role, namely to incorporate the lattice ${\cal
L}_{\epsilon=0}$ which is difficult to handle by the evolution method
described in Sec. \ref{s5.2}. Therefore our discussion will be brief.

\begin{figure}[tbh!]
\begin{center}
\includegraphics[width=5in,angle=0]{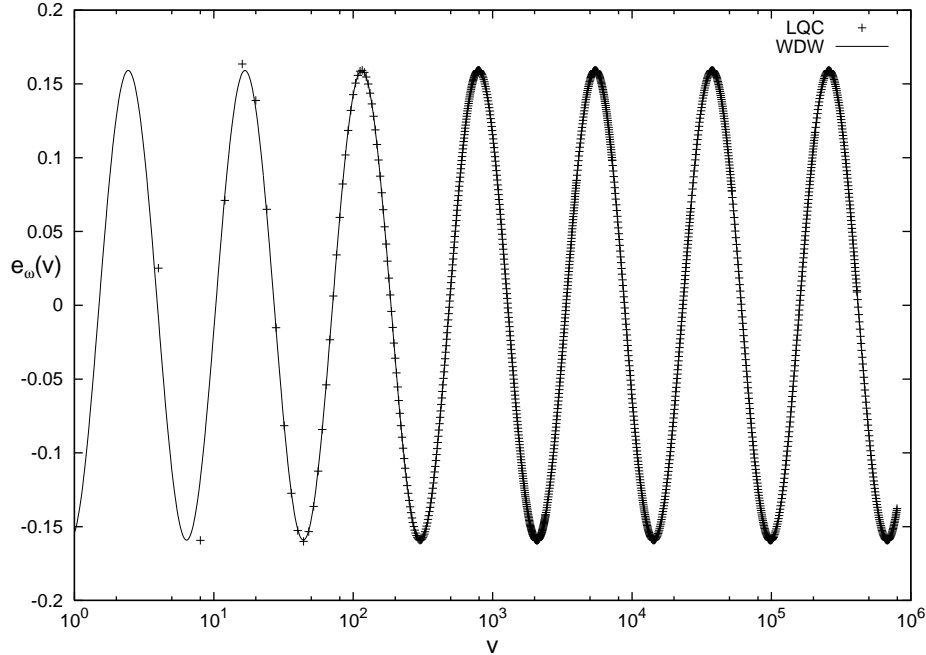}
\caption{Eigenfunction $e_{\omega}(v)$ of $\Theta$ (denoted by
$+$) for $\epsilon=0$ and $\omega=20$ is compared with
eigenfunction $\ub{e}_\omega(v)$ of the $\ul\Theta$ operator in
\WDW theory (solid line). At large $v$ both eigenfunctions
approach each other, however they differ at small $v$.}
\label{fig:e-LQC-WdW}
\end{center}
\end{figure}

Let us choose a lattice, say $\La_{|\epsilon|}$. Consider an
eigenfunction $\ub{e}_{\omega}(v)$ of
$\hat{C}_{\text{grav}}^{\text{WDW}}$, i.e. a solution to
$\ul{\Theta}\ub{e}_{\omega}(v) = \omega^2\ub{e}_{\omega}(v)$.
Since the left side of this equation approaches
$(\hat{C}_{\text{grav}}^{\WDW})
\ub{e}_{\omega}|_{\La_{|\epsilon|}}$ as $|v|\to\infty$ one expects
every solution $e_{\omega}(v)$ to \eqref{eq:eigen} to converge to
some \WDW eigenfunction $\ub{e}_{\omega}(v)$ in this limit. This
expectation was verified numerically, following the same procedure
as the one used in \cite{aps2}. Numerical simulations also
provided the rate of approach%
\footnote{The numerical tests have shown that the quantity
  $v^2|e_{\omega}(v)-\ub{e}_{\omega,\pm}(v)|$ is bounded.}:
for any $e_{\omega}(v)$ there exist eigenfunctions
$\ub{e}_{\omega,+}(v)$, $\ub{e}_{\omega,-}(v)$ of
$\hat{C}_{\text{grav}}^{\text{WDW}}$ (corresponding to the same
eigenvalue $\omega^2$) such that:
\begin{equation}\label{eq:e-asympt}
  e_{\omega}(v)\ = \begin{cases}
    \underline{e}_{\omega,\, +} (v) + O\left( \frac{1}{v^2} \right) \ ,
      & \text{for }v>0 \ , \\
    \underline{e}_{\omega,\, -} (v) + O\left( \frac{1}{v^2} \right) \ ,
      & \text{for }v<0 \ .
  \end{cases}
\end{equation}
An example of $e_{\omega}(v)$ and its approach to
$\ub{e}_{\omega,+}(v)$ for large positive $v$ is presented in fig.
\ref{fig:e-LQC-WdW}.

Each eigenfunction $\ub{e}_{\omega}(v)$ can be expressed as the
combination of basis functions $\ub{e}_k(v)$ defined via
Eq.(\ref{ek_eq}):
\begin{subequations}\label{eq:asympt-coeffs}\begin{align}
  e_{\omega}(v) &\xrightarrow{v\gg 1}
    A\,\ub{e}_{|k|}(v) + B\,\ub{e}_{-|k|}(v), &
  e_{\omega}(v) &\xrightarrow{v\ll -1}
    C\,\ub{e}_{|k|}(v) + D\, \ub{e}_{-|k|}(v)
  \tag{\ref{eq:asympt-coeffs}}
\end{align}\end{subequations}
Since the eigenfunctions $e_\omega(v)$ of $\Theta$ are determined
on the entire lattice $\La_{|\epsilon|}$ by their values on (at
most) two points, the WDW limits for positive and negative $v$ are
not independent. Thus, the coefficients $C,D$ are uniquely
determined by values of $A,B$ (and vice versa).

The property \eqref{eq:e-asympt} allows us to construct a basis $e_{k}^{(s)}(v)$ 
for the physical (symmetric) sector following step by step
the procedure presented in \cite{aps2}:
\begin{enumerate}
  \item One first constructs the basis functions $e^{\pm}_{-|k|}(v)$
(supported on $\La_{\pm|\epsilon|}$ respectively) by solving
\eqref{eq:eigen} on domains
$\La_{\pm|\epsilon|}\cap[-v^\star_{\pm},v^\star_{\pm}]$ for some
fixed, large, positive $v^\star_{\pm}\in \La_{\pm|\epsilon|}$. The
initial conditions are specified by demanding that the values of
$e^{\pm}_{-|k|}(v)$ agree with those of $\ub{e}^{\pm}(v)$ at $v =
v_{\pm}^\star, v_{\pm}^\star+4$. As in the case studied in
\cite{aps2}, eigenfunctions are enormously amplified on the
negative $v$ side and their limits for $v\ll -1$ are the
combinations $C\ub{e}_{|k|}+D\ub{e}_{-|k|}$ of WDW basis functions
with coefficients of almost equal absolute value i.e.,
$|C^{\pm}|\approx|D^{\pm}|$.
 \item Next, the functions $e^{\pm}_{-|k|}(v)$ are used to
calculate symmetric eigenfunctions $e^{(s)}_{-|k|}(v)$ via
\eqref{eq:e-symm}. Again their behavior is dominated by properties
of $e^{\pm}_{-|k|}(v)$ for $v<0$. In particular their limit for
large $v$ is composed of incoming $(k>0)$ and outgoing $(k<0)$ WDW
eigenfunctions. The coefficients of decomposition with respect to
the basis defined by \eqref{ek_eq} have almost equal absolute
value for $\epsilon \neq 0,2$:
    \begin{equation}\label{eq:phase-def}
      e^{(s)}_{-|k|}\mid_{\La_{\pm|\epsilon|}} \xrightarrow{v\to\infty}
      z^{\pm} ( e^{i\alpha^{\pm}}\ub{e}_{|k|} +
      e^{-i\alpha^{\pm}}\ub{e}_{-|k|} ) \ ,
    \end{equation}\label{eq:eks-rot}
where $z^{\pm}$ are some complex constants satisfying
$|z^+|\approx|z^-|$, while the phases $\alpha^{\pm}$ are functions
of $\epsilon$ and $\omega$. For $\epsilon = 0$ and $\epsilon = 2$,
$|z^+|=|z^-|$ and $\alpha^+ = \alpha^-$.

\end{enumerate}

These numerically constructed symmetric eigenfunctions can be now
used to build the semiclassical states, using \eqref{eq:psi-int}.
Our aim is to construct a state sharply peaked at a phase space
point lying on a classical trajectory of the expanding universe at
late times. The form of the physical inner product and the
expression of general physical state given by \eqref{eq:psi-int}
provides a natural choice of $\tilde{\Psi}(k)$:
\begin{equation}\label{eq:psi-k}
  \tilde{\Psi}(k) := e^{-\frac{(k-k^\star)^2}{2\sigma^2}}e^{-i\omega\phi^\star}
\end{equation}
where $\sigma$ is a suitable small spread. The resulting physical
state $\Psi(v,\phi)$ is peaked at $p_{\phi} = p_{\phi}^\star =
-\sqrt{12\pi G\hbar^2}k$. The value $v^\star$ of the Dirac
observable $\widehat{|v|_{\phi_o}}$ at which the state will be
peaked at $\phi=\phi_o$ is determined by $\phi^\star$. To obtain a
semiclassical state at late times we have to choose large value of
$p_{\phi}: p_{\phi}^\star\gg\hbar$ (in units c=G=1). Thus we need
to select $k^\star\ll -1$. The resulting $\tilde{\Psi}(k)$ has
such a small amplitude for $k > 0$ that without loss of numerical
precision it can there be set to zero. Therefore the explicit form
of $e^s_{|k|}$ is not needed.

Using expression  \eqref{eq:psi-k} in \eqref{eq:psi-int} we obtain
final form of the wave function:
\begin{equation}\label{eq:state-direct}
  \Psi(v, \phi) \ =\ \int_0^{\infty} \, d k \,
    e^{-\f{(k - k^\star)^2}{2 \sigma^2}} \, e^{(s)}_{k}(v)\,
    e^{i \omega(k)(\phi - \phi^\star)} \ .
\end{equation}
In order to calculate the integral we select a set of $k's$
uniformly distributed within the interval
$[k^\star-10\sigma,k^\star+10\sigma]$, compute the eigenfunctions
$e^s_{-|k|}(v)$ for all $k$ in this set and in the chosen finite
domain in $v$, and finally evaluate the integral
\eqref{eq:state-direct} using Fast Fourier Transform. Numerical
simulations were performed with various values of $p_\phi^\star$
ranging between $100$ and $1000$. Typical number of points
constituting the interval $k$ ranged from $2048$ to $4096$.

\subsection{Evolution}
\label{s5.2}

As we discussed in Sec. \ref{s4}, the quantum constraint equation  can
be cast in the form of a Klein-Gordon equation in a static spacetime
with $\phi$ playing the role of time. It is then natural to consider
the evolution in terms of the initial value problem in $\phi$. Since
$\Theta$ is a discrete operator in $v$, the quantum evolution reduces
to solving a system of coupled, ordinary, 2nd order differential equations.

To solve the quantum constraint and determine the evolution of the
state $\Psi(v,\phi)$ numerically, we have to first specify the
domain of integration. It was chosen to be $|v - \ep| \leq 4 N $
where $1 \ll N \in \mathbb{N}$ is an integer, such that the
initial (and, as the evolution shows, also subsequent) wave
function is negligibly small at the boundary. Nonetheless, since
this domain is finite we have specify boundary conditions. We
recall that for $|v| \gg 1$ the difference equation can be very
well approximated by the WDW equation. In particular the
fundamental equation $i\partial_\phi \Psi = \sqrt{\Theta} \Psi$
can be approximated by $\partial_\phi \Psi = (\sqrt{12 \pi G}) v
\partial_v \Psi$. To ensure deterministic evolution, the wave
packets $\Psi(v,\phi)$ were required to leave (rather than enter)
the domain of integration. The resulting equation at the boundary
was again discretized, taking the form:
\be
\partial_\phi \Psi(v,\phi) = s \, \sqrt{3 \pi G/4} \left(|v| -
2\right) \, \left(\Psi(v,\phi) - \Psi(v - 4 \, \mathrm{sgn}(v), \phi)
\right) ~.
\ee

Returning to the problem of evolution on the entire domain the initial
data $\Psi|_{\phi_o}$, $\partial_{\phi}\Psi|_{\phi_o}$ at $\phi =
\phi_o$ were chosen to be the same as those of the WDW Gaussian
semiclassical state \eqref{sc} corresponding to a large expanding universe,
peaked at $p_{\phi}=p_{\phi}^\star$ and $v|_{\phi_o}=v^\star \gg 1$. Note
that eigenfunctions of the quantum constraint $e^s_{-|k|}(v)$ approach
a linear combination of eigenfunctions of WDW constraint for $|v| \gg 1$
(Eq.(\ref{eq:phase-def})). Thus in order to keep the properties of the
initial state specified at $|v| \gg 1$ as close as possible to those
of the solution $\Psi(v,\phi)$ of the $\Theta$ operator at large
values of $|v|$, each of the basis functions $\ub{e}_{-|k|}$ was
rotated by a phase $e^{-i \alpha_\pm(k,\ep)}$.
The phase was found numerically by method
analogous to the one used in \cite{aps2}. After its values were
determined for variety of different $k$, the function of the
form
\begin{equation}
  \alpha^{+}(k) = A\ln(Bk+C)+D
\end{equation}
(where $A,B,C,D$ are real constants) was fitted to them. As the
0th and 1st order of its expansion in $k$ corresponds respectively
to a constant phase and the shift of origin in $\phi$ these terms
were subtracted from $\alpha^{+}$. The resulting WDW solution then
took the form:
\begin{equation}\label{eq:id}
  \ul{\Psi}(v,\phi) = \int dk \, e^{-\frac{(k-k^*)^2}{2\sigma^2}} \,
    e^{-i\alpha^{+}{}'} \, \ub{e}_{-|k|}(v) \,
    e^{i\omega(k)(\phi-\phi^\star)} \ ,
\end{equation}
where
\begin{subequations}\label{eq:id-sup}\begin{align}
  \alpha^{+}{}'(k) &:= \alpha^{+}(k)-\alpha^{+}(0)
    -k\partial_{k}\alpha^{+}(k)|_{0} \ , &
  \phi^\star &:= \phi_o - \sqrt{1/12\pi G}\ln|v^\star| \ .
    \tag{\ref{eq:id-sup}}
\end{align}\end{subequations}
The right side of \eqref{eq:id} and its derivative w.r.t $\phi$
were calculated numerically to obtain the desired initial data.

\begin{figure}[tbh!]
  \begin{center}
    \includegraphics[width=5in,angle=0]{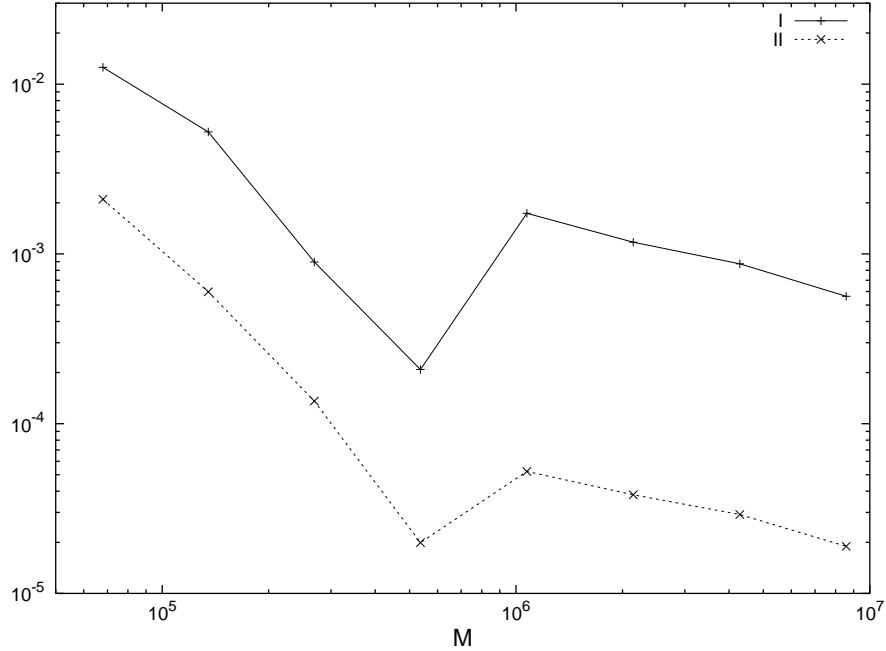}
    \caption{Error functions $\|\Psi_{(M)} - \Psi_{}\|/\|\Psi\|$
(upper curve) and $\||\Psi_{(M)}| - |\Psi_{}|\|/\|\Psi\|$ (lower
curve) are plotted as a function of time steps. Here
$\Psi_{(M)}$ refers to final profile of wave function for a
simulation with $M$ time steps. $\Psi_{}$ is the limit of final
profile as $1/M \rightarrow 0$ calculated via polynomial
extrapolation. In both cases, the evolution began at $\phi=0$ and
the final profile is evaluated at $\phi = -1.35$.}
\label{fig:conv-test}
\end{center}
\end{figure}

This initial data was then evolved using fourth order adaptive
Runge-Kutta method. To estimate the numerical errors due to
discretization in $\phi$ the profiles $\Psi|_{\phi}$ calculated
for different step sizes were compared. As the measure of the
`distance' between them we used the norm
\begin{equation}
\|f\|(\phi) \ := \ \sup_{|v_i-\epsilon|\leq N} |f(v_i,\phi)| \ .
\end{equation}
The step sizes were refined until the distance between the results
$\Psi_{\Delta\phi},\Psi_{\Delta\phi/2}$ of integration, with step
size $\Delta\phi$ and $\Delta\phi/2$ respectively, satisfied the
inequality
\begin{equation}
  \| \Psi_{\Delta\phi}-\Psi_{\Delta\phi/2} \|
  \leq \| \Psi_{\Delta\phi/2} \| \, (\Delta\phi) \,\, {\delta}
\end{equation}
for a pre-specified, small ${\delta}$. An example of behavior of
the distance for particular calculation is presented on fig.
\ref{fig:conv-test}. They show that the numerical errors manifest
themselves mainly in phases. The differences between the absolute
values of the wave function profiles are approximately one order
of magnitude smaller. Thus the expectation values and dispersions
of observables $\widehat{|v|_{\phi}}$ are determined with much
better precision than $\Psi$ itself.

%
%

Numerical simulations were performed using 10 different values of
$p_\phi^\star$  which ranged between $10^3$ and $2\times10^4$. The
value of $v$ at which the state peaks was always chosen to be
greater than $2.5 p_\phi^\star$. We chose 8 different $\epsilon$
sectors, with $\epsilon$ values uniformly distributed in the
interval between 0 an 2, excluding 0. The values of dispersions
$\sigma/k^\star$ varied from $1.7\%$ to $11\%$ depending on the value
of $p_{\phi}^\star$.

The resulting wave functions $\Psi(v,\phi)$ of the exact theory
were finally used to calculate the expectation values
$\left<\widehat{p_{\phi}}\right>$,
$\left<\widehat{|v|_{\phi}}\right>$ of observables defined by
\eqref{dirac4}. Using the inner product
$\left<\Psi|\Psi\right>_{\epsilon}$ given by \eqref{ip2}, their
explicit expressions are given by
\begin{subequations}\label{eq:num-expect}\begin{align}
  \left<\Psi\left|\widehat{|v|_{\phi}}\right|\Psi\right>
    &= \left<\Psi|\Psi\right>_{\epsilon}^{-1}
    \sum_{v\in\La_{\epsilon,N}}
    B(v) |v||\Psi(v,\phi)|^2\\
  \left<\Psi\left|\widehat{p_{\phi}}\right|\Psi\right>
    &= \left<\Psi|\Psi\right>_{\epsilon}^{-1}
    \sum_{v\in\La_{\epsilon,N}}
    B(v) \bar{\Psi}(v,\phi)(-i\hbar)\partial_{\phi}\Psi(v,\phi)
\end{align}\end{subequations}
where $\La_{\epsilon,N} := \{v=\pm\epsilon+4n;-N\leq n\leq N\}$

The dispersions
\begin{subequations}\label{eq:num-disp}\begin{align}
  \left<\Delta\widehat{p_{\phi}}\right>^2
    &= \left<\widehat{p_{\phi}^2}\right>
    - \left<\widehat{p_{\phi}}\right>^2  \\
  \left<\Delta\widehat{|v|_{\phi}}\right>
    &= \left<\widehat{v^2_{\phi}}\right>
    - \left<\widehat{|v|_{\phi}}\right>^2
\end{align}\end{subequations}
were also calculated.

\begin{figure}[tbh!]
  \begin{center}
    \includegraphics[width=5in,angle=0]{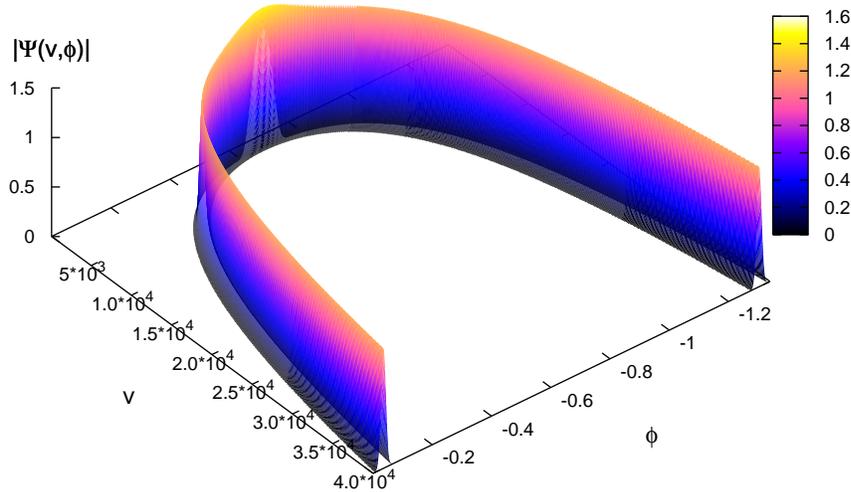}
    \caption{The absolute value of the wave function obtained
by evolving an initial data using the method of $\phi$ evolution.
For clarity of visualization, only the values of $|\Psi|$ greater
than $10^{-4}$ are shown. Being a physical state, $\Psi$ is
symmetric under $v \rightarrow -v$. In this simulation, the
parameters were: $\epsilon=2$, $p_{\phi}^{\star}=5 \times 10^3$,
and $\Delta p_\phi/p_\phi^{\star} = 0.025$.}
    \label{fig:l-3d}
  \end{center}
\end{figure}

\begin{figure}[tbh!]
  \begin{center}
    \includegraphics[width=5in,angle=0]{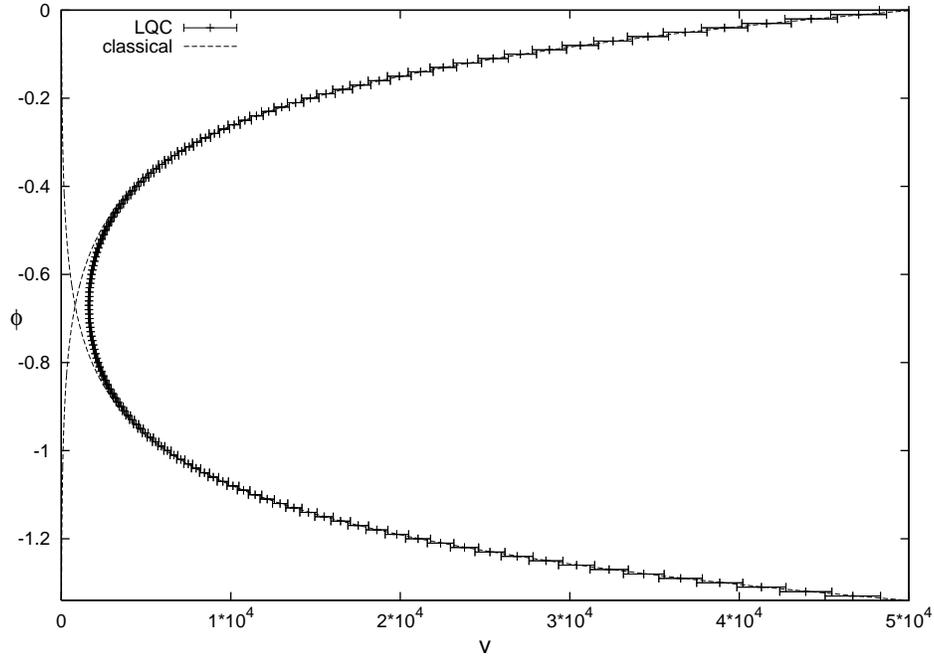}
    \caption{The expectation values (and dispersions) of
     $\widehat{|v|_{\phi}}$ are plotted for the wave function in
     Fig. \ref{fig:l-3d} and compared with expanding and contracting
     classical trajectories.}
    \label{fig:l-v}
  \end{center}
\end{figure}

\begin{figure}[tbh!]
  \begin{center}
    \includegraphics[width=5in,angle=0]{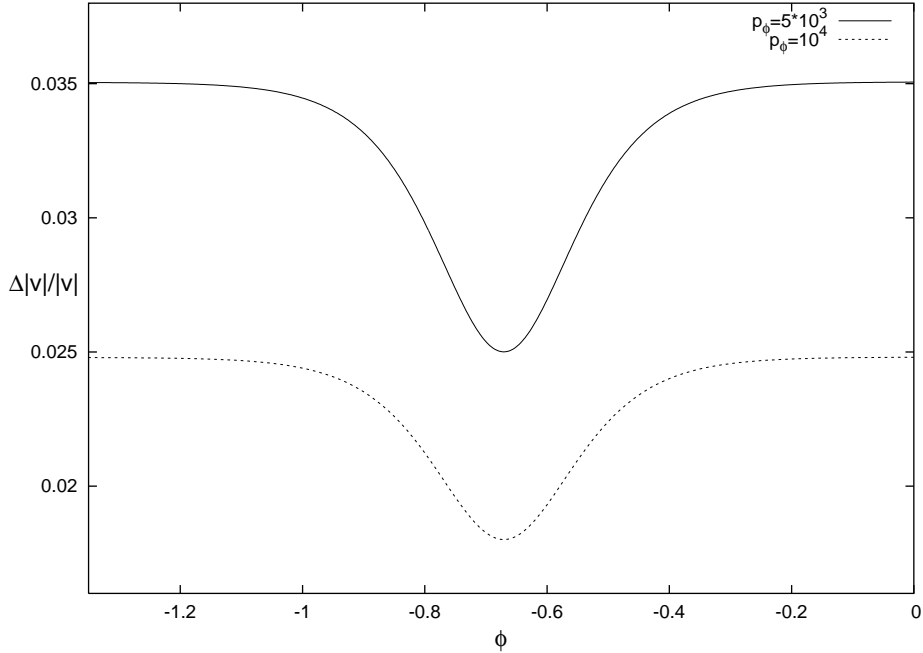}
    \caption{Plot of the behavior of relative dispersion
$\Delta |v|/|v|$ as function of $\phi$. The upper and lower curves
correspond to $\Delta p_\phi/p_\phi^{\star} = 0.025$ and $\Delta
p_\phi/p_\phi^{\star} = 0.018$ respectively. As can be seen
$\Delta|v|/|v|$ is asymptotically a constant for both expanding
and contracting branches.  }
    \label{fig:rel-dv}
  \end{center}
\end{figure}

\begin{figure}[tbh!]
  \begin{center}
    \includegraphics[width=5in,angle=0]{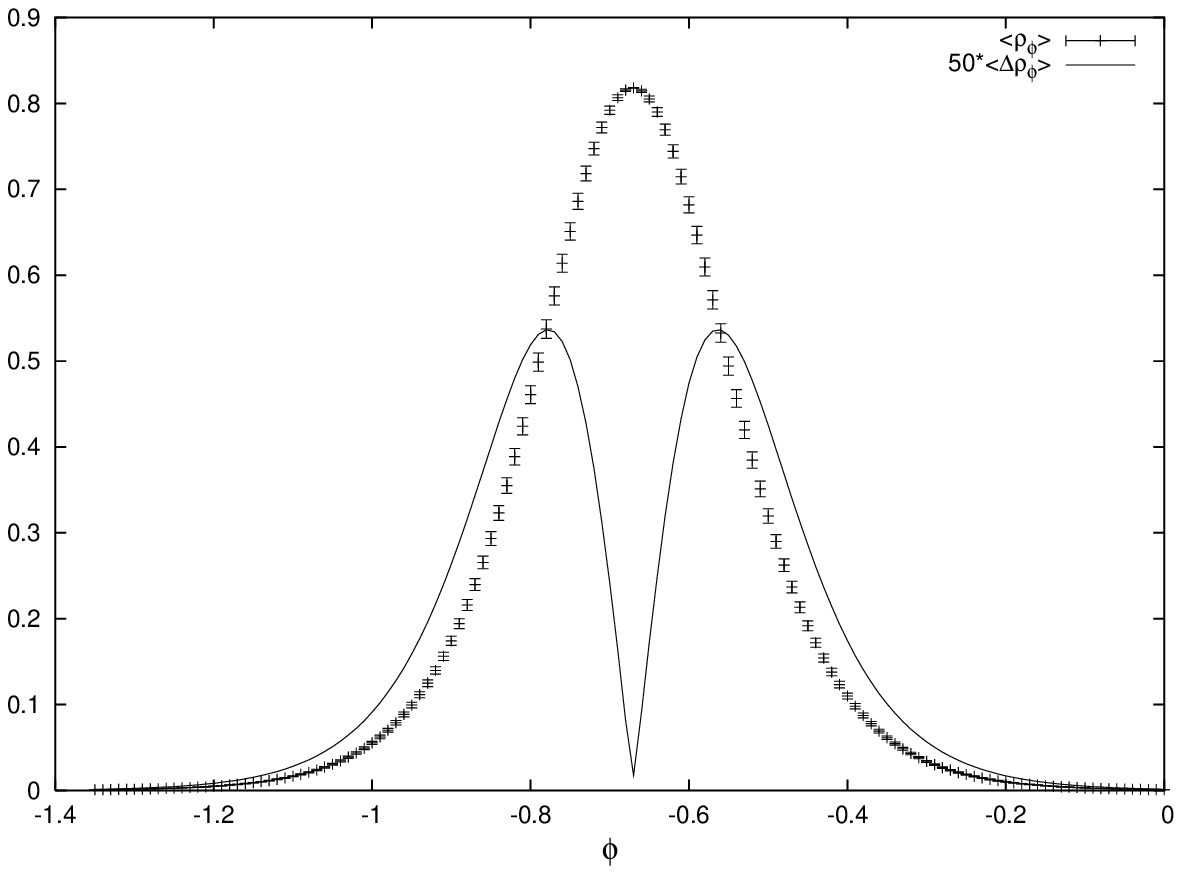}
    \caption{Plot of the behavior of $\langle\rho_\phi\rangle$ (points
with error bars) and its absolute dispersion (which for clarity is
multiplied by 50 and is shown by solid curve). Value of
$p_\phi^{\star} = 10,000$ and $\Delta p_\phi/p_\phi^{\star} =
0.018$.}
    \label{fig:rho-disp}
  \end{center}
\end{figure}

\subsection{Results}
\label{s5.3}

Results of various numerical simulations can be summarized as
follows:

\begin{enumerate}[ (1)]
\item The states remain sharply peaked throughout the evolution.
Their norms are preserved under the `$\phi$-evolution', providing
a consistency check on the numerics.

\item The classical trajectory provides a good approximation to
the expectation values of Dirac observables when the energy density
$\rho$ of the scalar field is small compared to a critical energy
density $\rcr$. However, when this value is reached, the
expectation values cease to be peaked at the classical solution.
Instead of following the classical trajectory into the big-bang
singularity as in the case of the \WDW dynamics, the LQC state
undergoes a quantum bounce. The numerical value of $\rcr$ was the
same in all simulations, given by $\rcr \approx 0.82\rho_{\rm Pl}
\equiv 0.82/G^2\hbar$. A physical understanding of this phenomenon
can be obtained from the modified Friedmann equations which
incorporate the leading corrections due to quantum geometry
effects (see Appendix \ref{a2.1}). These effective equations also
provide an analytical expression $\rcr = 3/(16 \pi^2 \gamma^3 G^2
\hbar)$ of the critical density whose numerical value agrees with
that found in the simulations of the exact LQC dynamics.

\item On further backward evolution $\rho$ decreases. When it
becomes small compared to $\rcr$, the state again becomes sharply
peaked on a classical solution which has the same value of
$\langle \hat p_\phi \rangle$, but which is contracting in the
future. Thus, the quantum geometry effects lead to a resolution of
the big bang singularity. Furthermore, the state becomes
semi-classical also in the distant past, and the pre and post
big-bang branches are joined by a `quantum bridge' by the
deterministic evolution of LQC. So long as the initial state at
late times is chosen to be semi-classical, this scenario,
including the value of $\rcr$, is robust.
%
%
An example of the result of numerical simulation is shown in Fig.
\ref{fig:l-3d}. The comparison between the classical trajectories
and quantum evolution is presented in Fig. \ref{fig:l-v}.

\item In order to better understand the behavior of energy density
during quantum evolution we \emph{independently} analyzed the
evolution of expectation values of the energy density operator,
defined as $\hat \rho_\phi = \widehat{p_\phi^2/2 \, p^3|_\phi}$.
Results were consistent with those reported above. We found that
in all quantum solutions $\langle \hat \rho_\phi \rangle$ is
bounded above by $\rcr \approx 0.82 \rho_{\mathrm{Pl}}$. An
example of the behavior of the expectation values of $\hat
\rho_\phi$ is presented in Fig. \ref{fig:rho-disp}.

\item In the regime where quantum gravity modifications to the
classical dynamics are negligible, the relative dispersions of
Dirac observables were found to remain constant and {\it equal} on
both sides of the bounce point (see Fig. \ref{fig:rel-dv}). Thus
both semiclassical branches, before and after the bounce, are
equally coherent. Close to the bounce point the value of relative
dispersion of $\widehat{|v|_{\phi}}$ decreases reaching minimal
value exactly at the bounce point. However, this decrease does not
imply increased coherence in the large density region. Indeed
using the property that states remain sharply peaked, we can find
$\Delta \phi$ from $\Delta v/v$ and estimate the uncertainty
product $\Delta \phi \Delta p_\phi$. While the product remains
roughly constant before and after the bounce, it has a small
\emph{increase} near the bounce point.

Behavior of the dispersion $\Delta \rho_\phi$ is shown in Fig.
\ref{fig:rho-disp}. Its value grows in the classical regime in the
expanding phase but becomes very small near the bounce point.
After the bounce, dispersion again increases, before decreasing in
the classical contracting phase. This peculiar variation in
$\Delta \rho_\phi$ near the quantum bounce can be qualitatively
understood using effective theory discussed  in Appendix B. It has its origin
in the phenomena of super-inflation --- i.e. the phase in which
$\dot H > 0$ where $H \equiv \dot a/a$ is the Hubble rate--- which
occurs in the regime $\rcr > \rho > \rcr/2$ \cite{singh:2006a}.
However, since some of the assumptions that underlie the current
derivation of the effective equations are generally violated in
the Planck regime, this `unreasonable efficacy' of the effective
equations remains somewhat mysterious.

\end{enumerate}

\section{Discussion}
\label{s6}

This is the second in a series of detailed papers whose goal is to
develop a comprehensive LQG framework to systematically
investigate the physics of the Planck regime near the big bang. In
\cite{aps2} we analyzed the homogeneous isotropic model with a
massless scalar field by introducing several new techniques to
construct the physical sector of the theory. We established that,
thanks to the quantum geometry effects which distinguish LQC from
the \WDW theory, in the backward evolution of states which are
semi-classical at late times, the big bang is replaced by a
quantum bounce. That investigation was based on the quantum
Hamiltonian constraint that has been used in the LQC literature
for the last three years. We were able to analyze the physical
sector of the theory in detail and firmly establish that, although
the quantum evolution dictated by that constraint has several
desirable features, it also has a serious flaw: quantum effects
can dominate and significantly modify classical predictions even
when the matter density (and curvatures) are low \cite{aps2,dh2}.
In this paper we showed that a physically motivated modification
of the quantum Hamiltonian constraint overcomes this weakness
\emph{while preserving the attractive features of the older
evolution.} Indeed, a key strength of this work is that several
features of the present simulations are qualitatively similar to
those of \cite{aps2}. Finally, in both works, detailed numerical
analysis of dynamics was restricted to those quantum states which
are semi-classical at late times. Within the model, these states
best represent our physical universe and our discussion of the
quantum bounce refers only to these states, although evolution is
well defined and unitary on the full physical Hilbert space.

We deliberately followed the same organization as \cite{aps2} to
bring out the (similarities and) differences between the two
quantum evolutions. Physically the differences are critically
important but mathematically they are rather subtle. This is just
as one would expect of an improvement that cures a significant
limitation while retaining the strengths of the older method. In
Appendix A we outline the inclusion of the cosmological constant.
Again, while the previous quantum evolution shows certain
departures from the classical theory even when the space-time
curvature is low \cite{dh2}, the quantum evolution generated by
the new Hamiltonian constraint is free of this drawback. Now the
departures occur only in the deep Planck regime near the big bang
(or the big crunch) and lead to a replacement of the classical
singularity by a quantum bounce.

Models discussed so far are too simple to be physically realistic.
However, the methods introduced are rather general. In particular,
as indicated at the end of section \ref{s2.1}, the
`$\bar\mu$-strategy' incorporates certain key features of full
LQG. Hence, results obtained in this paper provide concrete
indications of how quantum geometry can lead to subtle yet
important departures from the classical theory by making gravity
repulsive in the deep Planck regime. Glimpses of new physics that
may emerge were presented in the last section of \cite{aps2}. We
direct the reader to that paper also for a detailed discussion of
numerical analysis, physical ramifications of results and
comparisons with other approaches that also feature emergent time
and/or quantum bounce.

Finally, as emphasized in section \ref{s1}, a key limitation of
the present work is that the theory is not obtained through a
systematic symmetry reduction of full LQG. This is inevitable at
present because our understanding of the full quantum dynamics of
LQG is still very incomplete. The overall framework \emph{is}
constructed by following strategies developed in full LQG.
However, an important difference arose in the introduction of the
Hamiltonian constraint operator because our mini-superspace
reduction is carried out by gauge fixing and is therefore not
diffeomorphism invariant. As a result, in the last step we had to
use some physical considerations and borrow the expression of
the area gap $\Delta$ from the full theory.  The strategy in LQC
is to begin with simple models and work one's way up by
incorporating the lessons learned along the way. Improved dynamics
discussed in this paper is an interesting example of such lessons.
Considerable work is now in progress, also by several others, to
extend the present analysis to more complicated models that will
include anisotropies as well as inhomogeneities.

\bigskip
\textbf{Acknowledgments:} We would like to thank Martin Bojowald,
Frank Herrmann and especially Kevin Vandersloot for discussions.
This work was supported in part by the NSF grants PHY-0354932 and
PHY-0456913, the Alexander von Humboldt Foundation, the Krammers
Chair program of the University of Utrecht and the Eberly research
funds of Penn State.

\begin{appendix}

\section{Non-zero cosmological constant}
\label{a1}

In this appendix we will investigate the dynamics of the universe
with a massless scalar field and a cosmological constant
$\Lambda$. Our goal is only to test whether the quantum bounce
scenario is robust and if the $\bar\mu$-evolution is free of the
undesirable features of the $\mu_o$-evolution, such as departures
from the classical theory at low values of total energy densities
\cite{dh2}. Therefore the numerical simulations are not as refined
as in the main body of the paper. A more systematic and detailed
analysis will appear elsewhere.

A negative $\Lambda$ leads to a classical recollapse of the
universe when total energy density, including the `vacuum energy
density' due to $\Lambda$, vanishes. If a quantum bounce exists in
this model then it can lead to a cyclic model of the universe. A
positive $\Lambda$ is favored by current observations. For
completeness we will discuss both the cases and ask whether
general features of the $\Lambda=0$ analysis persist.

The classical Hamiltonian constraint is of the form
\begin{equation}
  C_{\Lambda}\ =\ -\frac{6}{\gamma^2}c^2\sqrt{p}
    + 2 \, \Lambda \, |p|^{\frac{3}{2}}
    + 8\pi G |p|^{-\frac{3}{2}} p_{\phi}^2 \ .
\end{equation}
From Hamilton's equations it is easy to see that  the momentum
$p_{\phi}$ is again a constant of motion and the scalar field
$\phi$ is a monotonic function of time.
%
%
Hence it is well suited to play the role of emergent time in the
quantum theory.

The quantization procedure and the construction of the physical
Hilbert space (and observables) for the case $\Lambda \neq 0$ is
completely analogous to that in the model with $\Lambda=0$. The
quantum Hamiltonian constraint it gives is similar in its form to
\eqref{qh7}
\begin{equation}\label{eq:evol-L}
  \partial_{\phi}^2 \Psi(v,\phi)
  \ =\ -\Theta_{\Lambda} \Psi(v,\phi)
  \ :=\ - \left[\Theta
                 - \frac{16  \pi^2  \gamma^3  \lp^4}{27  K  \hbar} [B(v)]^{-1} \Lambda v\right]\Psi(v,\phi) \ ,
\end{equation}
where $\Theta_{\Lambda}$ is a self-adjoint operator and $K$ is a
constant defined in \eqref{v}. For $\Lambda>0$  it fails to be
positive definite on $\mathcal{H}_{\text{kin}}$. However, since
the operator $\p_\phi^2$ on the left side of (\ref{eq:evol-L}) is
negative definite, only the projection of $\Theta_{\Lambda}$ on
its positive eigenspace is relevant for solutions to the
constraint. We can therefore repeat the procedure of section
\ref{s4.2}, decompose the solutions into positive and negative
frequency parts, and construct the physical inner product and
observables. The expectation values of $\widehat{|v|_{\phi}}$,
$\widehat{p_{\phi}}$ and their dispersions at an instant $\phi$
are again given by \eqref{eq:num-expect} and \eqref{eq:num-disp}
respectively, where as the norm of the wave function can be
calculated via \eqref{ip2}.

The next step is to construct states which are sharply peaked at
late times and compare the behavior of the expectation values of
observables with classical trajectories. The particular properties
of the model differ for $\Lambda<0$ and $\Lambda>0$. Therefore
they have to be handled with different methods and we do so
separately in sections \ref{a1.1} ($\Lambda<0$) and \ref{a1.2}
($\Lambda>0$).

\subsection{Negative Cosmological Constant}
\label{a1.1}

The classical equations of motion imply that $v(\phi)$ satisfies
the following differential equation:
\begin{equation}\label{eq:vL-ode}
  \left( \frac{\partial_{\phi}v}{v} \right)^2\
  =\ 12\pi G
     + \frac{(4\pi\gamma)^3}{(3K)^2}v^2\frac{\Lambda}{p^2_{\phi}}
     \ .
\end{equation}
When the cosmological constant is negative the solution to this
equation takes the form
\begin{equation}
  v(\phi)\ =\ \sqrt{12 \pi G} \frac{3 K}{(4  \pi  \gamma  \lp^2)^{3/2}} \frac{p_\phi}{\sqrt{|\Lambda|}} |\cosh(\sqrt{12 \pi G}(\phi - \phi_o))^{-1}| \ .
\end{equation}
Thus, the universe originates at the big bang singularity (for
$\phi=-\infty$), expands until the recollapse point (at
$\phi=\phi_o$) where the total energy density (due to the scalar
field \emph{and} the cosmological constant) drops to zero and then
contracts, reaching the big crunch singularity at $\phi=+\infty$.

To determine the quantum evolution we apply the method specified in
section \ref{s5.2}, that is:
\begin{enumerate}[ (i)]
  \item We first choose on the $v$ axis the domain $\La_{\epsilon}\cap
    [-v_b,v_b]$, where $\La_{\epsilon}$ is a lattice defined in
    section \ref{s4.1} for some $\epsilon\neq 0$, and $v_b \gg 1$.
  \item Next, we specify the initial data, $\Psi(v,\phi_o)$,
  $\partial_{\phi}\Psi (v, \phi)|_{\phi_o}$, peaked at a phase space point
  representing an expanding, large classical universe.
  \item We evolve this data backward, solving \eqref{eq:evol-L} numerically
   as in section \ref{s5.2}.
  \item Finally we calculate the expectation values and their
    dispersions using \eqref{eq:num-expect} and \eqref{eq:num-disp}
\end{enumerate}

\begin{figure}[tbh!]
  \begin{center}
    \includegraphics[width=5in,angle=0]{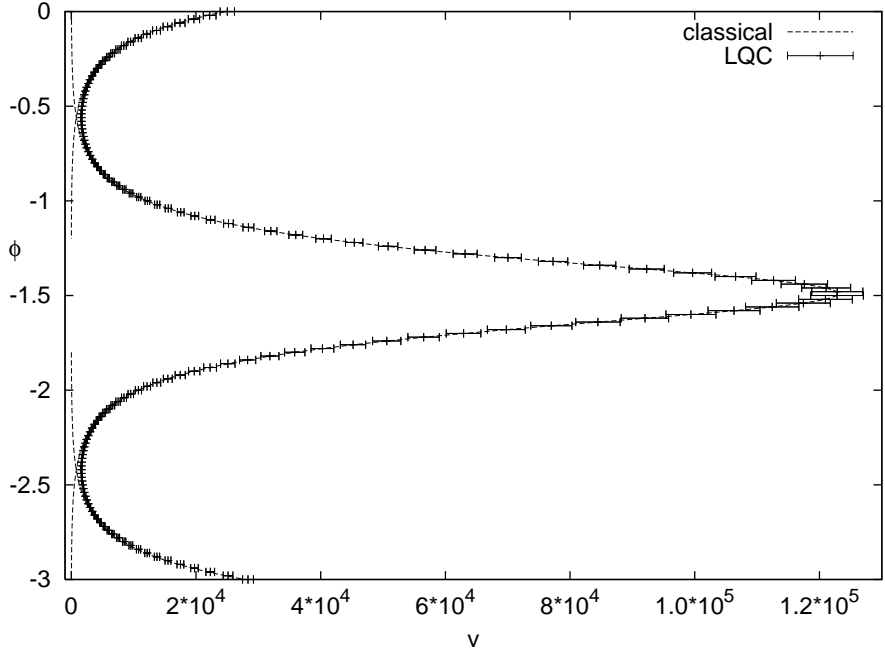}
    \caption{The expectation values (and dispersions) of
      $\widehat{|v|_{\phi}}$ for the model with negative cosmological
      constant are compared with classical trajectories. In this
      simulation, the parameters were: $\epsilon=2$, $p_{\phi}^{\star}
      =5 \times 10^3$, and $\Lambda = -3.6 \times 10^{-3}$,
      $\Delta p_\phi/p_\phi^{\star} = 0.034$.}
    \label{fig:negLambda}
\end{center}
\end{figure}

We specified the initial state by choosing a Gaussian in $(\mu,c)$
peaked at $\mu^\star = K^{-\frac{2}{3}} (v^\star)^{\frac{2}{3}}$ (see
\eqref{v}) and $c^\star(\mu^\star,p^\star_{\phi})$, and its time derivative by
using the classical trajectory. (Thus, we follow `method I' of
section V.B.2 of \cite{aps2}). As discussed there, while this
method is not as optimal as the one used in main body of this
paper, it has the advantage that it does not require the knowledge
about properties of the eigenfuncions of $\Theta_{\Lambda}$. The
exact forms of these initial profiles were the following:
\begin{subequations}\begin{align}
  \Psi(v,\phi)|_{\phi_o} &= |v|^{\frac{1}{2}}
    e^{-\frac{(\mu(v)-\mu^\star)^2}{2\sigma^2}}
    e^{-\frac{i}{2}(\mu(v)-\mu^\star)c^\star} \\
  \partial_{\phi} \Psi(v,\phi)|_{\phi_o} &= \ \Psi(v,\phi_o) \frac{\gamma  \beta^2}{2  p_\phi} \left[\frac{i  \mu(v)}{3} \left(- \Lambda  {\mu^*}^2  + \frac{8  \pi  G  p_\phi^2}{\beta^3  \mu^*}\right) + 
i {\mu^*}^3 \Lambda + \frac{4 (\mu(v) - {\mu^*}) {\mu^*}^2  c^*}{\beta \gamma^2 \sigma^2} \right] \ .
\end{align}\end{subequations}
where $\beta:= 4\pi\gamma\lp^2/3$.

Because the domain of integration in numerical simulations is
compact in $v$, we have to provide the boundary conditions. Unlike
for $\Lambda=0$ however the size of classical universe is bounded
by a maximum value $v_{\rm max}$. Therefore, it suffices to choose
$v_b$ such that the classical recollapse point is well within the
domain and set the `mirror boundary conditions': $\Psi =
\partial_{\phi}\Psi = 0$. In numerical simulations $v_b$ was
chosen to satisfy the inequality $v_b\geq 1.2\, v_{\rm max}$ to
make sure that the recollapse occurs due to dynamics and is not an
artifact of the boundary conditions.

An example of the result of simulation is presented in fig.
\ref{fig:negLambda}. The state remains sharply peaked throughout
the evolution. The expectation values follow classical trajectory
corresponding to the expansion phase till the {\emph total} energy
density
\begin{equation}\label{eq:rho-tot}
  \rho_{\text{tot}} = \rho_{\phi} + \rho_\Lambda, \quad{\rm
  where}\quad \rho_\Lambda := (8\pi G)^{-1}\Lambda\, ,
\end{equation}
becomes comparable to $\rcr$. At $\rho_{\text{tot}}=\rcr$ the
universe bounces due to repulsive effects of quantum geometry, and
follows contracting phase of classical trajectory. The agreement
with classical trajectory remains good through the classical
recollapse and expansion phase preceding it, until the energy
density becomes large again. Then again a bounce is observed. In
effect the evolution is periodic: each `cycle' of classical
evolution is connected through quantum bridge with previous/next
cycle. Both the Big Bang and Big Crunch singularities are resolved
and replaced by quantum bounces. The behavior of fluctuations of
Dirac observables is also periodic in the sense that the spread at
a given point of the trajectory and after full cycle are the same.
Thus, in the fully quantum evolution of LQC, semi-classical states
do not loose coherence in evolution from one cycle to another.

\subsection{Positive Cosmological Constant}
\label{a1.2}

In the case when cosmological constant is positive, classical
trajectories, i.e., solutions to \eqref{eq:vL-ode}, take the form
\begin{equation}
  v(\phi)\ =\ \sqrt{12 \pi G} \frac{3 K}{(4\pi\gamma\lp^2)^{3/2}} \frac{p_\phi}{\sqrt{|\Lambda|}} |\sinh(\sqrt{12 \pi G}(\phi - \phi_o))^{-1}| \ .
\end{equation}
As for $\Lambda=0$ we have then two types of trajectories. In one
the universe starts from a big bang singularity, expands and
reaches infinite value of $v$ for finite $\phi=\phi_o$. In the
other the universe contracts from the infinite volume state (at a
finite $\phi=\phi_o$) and reaches the big crunch singularity.

The equation \eqref{eq:vL-ode} implies in particular that in the
region where energy density of scalar field is small with respect
to vacuum energy density $\rho_{\Lambda}$,
the speed $\partial_{\phi}v$ of a wave packet following classical
trajectory is proportional to $v^2$. This feature, and the fact
that to specify proper boundary conditions for evolution in $\phi$
we need to know an explicit form of the square root of the
Wheeler-DeWitt limit of the positive part of $\Theta_{\Lambda}$,
makes the application of the evolution method numerically
difficult. Therefore for the construction of semiclassical states
we used the method of direct evaluation of the integral
representation of $\Psi(v,\phi)$ specified in section \ref{s5.1}.
Thus,
\begin{enumerate}[ (i)]
  \item We first calculate the symmetric eigenfunctions $e_k^s(v)$ of
    the $|\Theta_{\Lambda}|$ operator.
  \item Next, we construct the Gaussian state peaked at some $k^\star$ and
    with spread $\sigma$ (see Eq. \eqref{eq:state-direct}).
  \item Finally, we evaluate the integral \eqref{eq:state-direct} using
    fast Fourier transform and calculate the expectation values.
\end{enumerate}
The exact application of this method requires the normalization of
$e^s_k(v)$ with respect to the inner product given by \eqref{ip2}.
However, our goal here is only to test robustness of the quantum
bounce in presence of a cosmological constant. Therefore, we will
not construct semi-classical states which minimize uncertainties
but work just with states which are reasonably sharply peaked. Then, it is
enough to use for basis eigenfunctions $e'{}^s_k(v) := C(k)
e^s_k(v)$ where $C(k)$ changes slowly with $k$, remaining
approximately constant within interval
$[k^\star-3\sigma,k^\star+3\sigma]$. To construct such functions
we first observe that, for large $v$, $e^s_k(v)$ show oscillatory
behavior and the absolute values of their extremas decrease as
$|v|$ increases and set the `normalization' by requiring equality
of maximas closest to the large $v^\star$ chosen initially.

\begin{figure}[tbh!]
  \begin{center}
    \includegraphics[width=5in,angle=0]{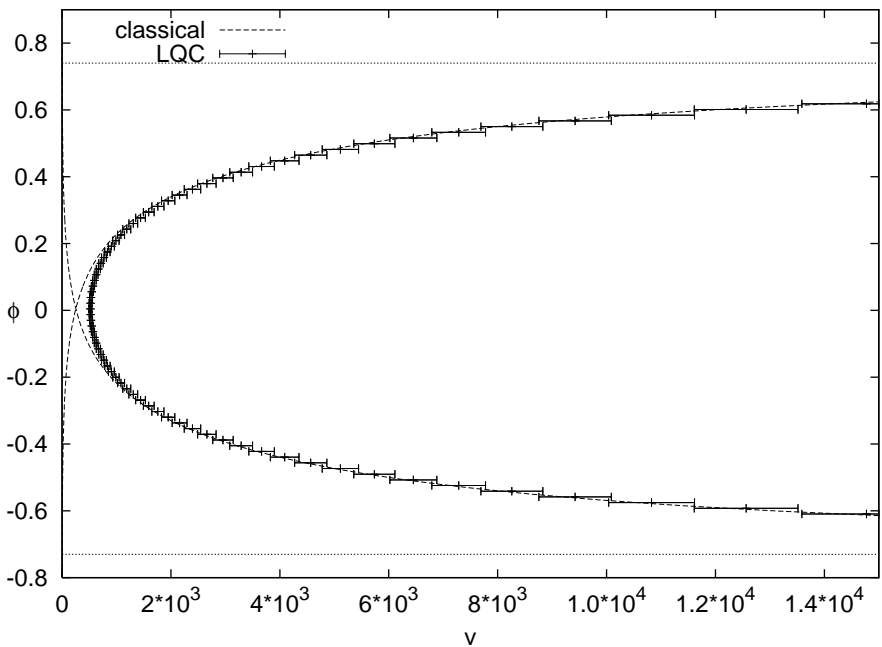}
    \caption{The expectation values (and dispersions) of
      $\widehat{|v|_{\phi}}$ for the model with positive cosmological
      constant are compared with classical trajectories. In this
      simulation, the parameters were: $\epsilon=2$, $p_{\phi}^{\star}
      =1.6 \times 10^3$, and $\Lambda = 4.1 \times 10^{-2}$. Horizontal
      dashed lines denote  asymptotic values of $\phi$ for which $v\to\infty$.}
    \label{fig:posLambda}
\end{center}
\end{figure}

An example of the result of simulation is presented in figure
\ref{fig:posLambda}. The state indeed is sharply peaked. As in the
case $\Lambda=0$, the expectation values of $\widehat{|v|_{\phi}}$
again reproduce the picture of two semiclassical  regions,
one contracting and the other expanding, connected by a quantum
bounce. The total energy density (defined via \eqref{eq:rho-tot}) at
the bounce point equals $\rcr$. Unlike in the $\Lambda=0$ case
however the universe shrinks from $v=\infty$, bounces and again
re-expands to $v=\infty$, all within a  compact interval of
$\phi$.

To summarize, in both cases ($\Lambda>0$ and $\Lambda<0$):
\begin{enumerate}
  \item Classical singularity is replaced by a quantum bounce, which
occurs when $\rho_{\text{tot}} = \rho_\phi + \rho_\Lambda = \rcr$.
The critical value of energy density was found numerically to be
equal to $0.82 \rho_{\mathrm{Pl}}$. Thus, \emph{the value of $\rcr$
is independent of the value of cosmological constant.} Analysis of
the modified Friedmann equation on the lines of Appendix B yields
the same expression, $\rcr = \sqrt{3}/(16 \pi^2 \gamma^3 G^2
\hbar)$, of the critical density.

  \item In the dynamics dictated by the Hamiltonian constraint used
so far in literature, a deviation from the classical behavior
resulting in a recollapse occurs at \emph{low} energy density
scales, when cosmological constant dominates \cite{dh2}. The
improved dynamics presented in this work is free of this
physically undesirable feature. More generally, the new dynamics
reproduces the standard Friedmann dynamics when $\rho \ll \rcr$.
\end{enumerate}

\section{Some conceptual issues}
\label{a2}

\subsection{Effective dynamics}
\label{a2.1}

An effective description of quantum dynamics can be obtained by
applying geometric methods to quantum mechanics, where the Hilbert
space is treated as an infinite dimensional phase space which
has a structure of a fiber bundle (see, e.g., \cite{as,jw}). The base
space of this bundle is the classical phase space and fibers are
the states with same expectation values of the operators for the
corresponding canonically conjugate phase space variables. Using
coherent states we can attempt to find horizontal sections which
are preserved under the \emph{quantum} evolution up to some
desired accuracy, and can thus obtain an effective Hamiltonian
which incorporates the leading quantum corrections to the
classical dynamics.

Let us consider the $\Lambda=0$ case as in the main body of the
paper. Then, for LQC, this procedure leads to an effective
Hamiltonian with leading order terms as \cite{vt}:
\be\label{heff0} \heff =  \f{C_{\rm eff}}{16\pi G} = -
\f{3}{8 \pi G \gamma^2 \bar\mu^2}
 |p|^{\f{1}{2}}\,\sin^2(\bar\mu c) + \frac{1}{2} \,B(p) \, {p_\phi^2} ~.
\ee
Here $B(p)$ denotes the eigenvalue of $\widehat{1/|p|^{3/2}}$
operator given by (\ref{inversevol}). Modifications to the
dynamics due to behavior of $B(p)$ become significant for $|v|
\sim 1$.
\footnote{The value of $|v|$ at which effects of $B(p)$ become
important depends on the value of the quantization ambiguity
parameter $j$ which following theoretical considerations of Refs.
\cite{kv,ap}, has been set equal to half.}
For $|v| \gg 1$, $B(p)$ quickly approaches the classical value
$|p|^{-3/2}$, proportional to $1/|v|$ and the rate of approach is
given by
\be
B(p) = \left(\f{6}{8 \pi \gamma \lp^2}\right)^{3/2} \, \f{K}{|v|}
\left(1 + \f{5}{9} \, \f{1}{|v|^2} + O\left(\f{1}{|v|^4}\right)\right)
~.
\ee
Neglecting the higher order quantum corrections to $B(p)$ we can
obtain the Hamilton's equation for $v$ and $\phi$:
\be \label{dotv} \dot v = \{v,\,\heff\} = - \f{8 \pi \gamma
G}{3}\, \f{\partial\heff}{\partial c}  = \frac{2 |v|^{1/3}}{\gamma
\bar\mu K} \, \left(\f{8 \pi \gamma \lp^2}{6}\right)^{1/2} \, \snf
\, \csf~ .  \ee
and
\be\label{dotphi} \dot \phi = \{\phi, \heff\} =
\left(\f{8 \pi \gamma \lp^2}{6} \right)^{-3/2} \, K \, \f{p_\phi}{|v|}\, . \ee

\begin{figure}[tbh!]
  \begin{center}
    \includegraphics[width=5in,angle=0]{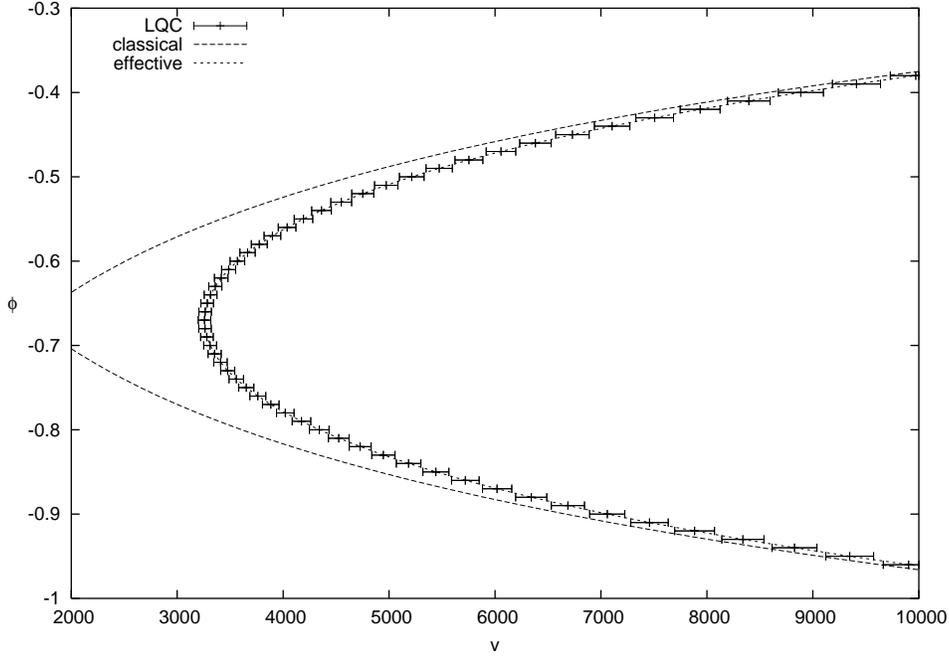}
    \caption{A comparison of the quantum dynamics represented by the
expectation values and dispersions of $\widehat{|v|_{\phi}}$ (points
with error bars), the
effective dynamics (dotted curve) and the classical dynamics (dashed
curve) is presented near the bounce point. The effective dynamics
provides a good approximation to the underlying quantum theory. The
values of parameters were $p_\phi = 10000$ and $\epsilon = 2$.}
    \label{fig:v-traj-zoom}
  \end{center}
\end{figure}

The modified Friedmann equation can be obtained from the vanishing of
the Hamiltonian constraint (\ref{heff0}):
\be \snfs = \frac{8 \pi  \gamma^2 \bar\mu^2 G}{6 \, |p|^2} \,
p_\phi^2\, , \ee and using Eq.(\ref{dotv}).
It turns out to be
\be \label{mod_fried} H^2 \equiv \f{\dot v^2}{9 v^2} = \frac{8 \pi
G}{3} \, \rho \left( 1 - \f{\rho}{\rcr} \right), \quad {\rm where}
\quad \rcr = \f{\sqrt{3}}{16 \pi^2 \gamma^3 G^2 \hbar} ~, \ee
with $\rho = p_\phi^2/2 |p|^3$. Quantum geometry effects thus lead
to $\rho^2$ modification of the Friedmann equation at the scales
when $\rho$ becomes comparable to $\rcr$. The Hubble parameter
vanishes when energy density becomes equal to the critical value
$\rcr$ and the universe bounces from the expanding (contracting)
branch to the contracting (expanding) branch. Further, $\rho/\rcr$
term becomes negligible for $\rho \ll \rcr$ and Eq.
(\ref{mod_fried}) reduces to the standard Friedmann equation;
there are no departures from classical general relativity in the
low curvature regime.
Detailed phenomenological investigations of the effective theory
obtained from the improved quantum constraint confirm that
unnatural effects of the old quantization are cured in the
improved quantization \cite{imp_pheno}.

To compare effective dynamics with the exact quantum evolution in
Sec. \ref{s5.2}, it is useful to combine (\ref{dotv}) and
(\ref{dotphi}) to obtain:
\be \label{mod_dmdf} \f{d v}{d \phi} = \sqrt{12 \pi G} \, \left(1
- \f{\rho}{\rcr}\right)^{1/2} \, v ~. \ee
Trajectories obtained from this equation are plotted in
Fig.\ref{fig:v-traj-zoom} and compared with the expectation values
of $\widehat{|v|_\phi}$ and the classical dynamics. As can be
seen, the effective dynamics provides an excellent approximation
to the underlying  quantum dynamics.

\textbf{Remark:} The modified field equations (\ref{mod_fried})
and (\ref{mod_dmdf}) can be interpreted as saying that the
effective Newton's constant is given by $G_{\rm eff} = G (1 -
{\rho}/{\rcr})$, where $G$ is the low energy Newton's constant and
$\rcr \approx 0.82\rho_{\rm Pl}$. Now, the renormalization group
analysis based on Euclidean quantum gravity \cite{mr} strongly
suggests the existence of a non-trivial fixed point at which the
theory becomes asymptotically free. The behavior of $G_{\rm eff}$
in LQC is in qualitative agreement with that picture.

\subsection{Subtleties associated with $k=0$ cosmologies}
\label{a2.2}

To write the space-time metric in the standard FRW form, $g_{ab} =
-\nabla_at \nabla_b t+ a^2(t) q^o_{ab}$, one needs a fiducial,
Riemannian 3-metric $q^o_{ab}$ of constant curvature $k$. Clearly,
under a rescaling $q^o_{ab} \rightarrow \alpha^2 q^o_{ab}$, the
scale factor $a(t)$ scales as $a(t) \rightarrow \alpha^{-1} a(t)$.
As is well-known, while this rescaling freedom can be eliminated
in the $k =\pm 1$ cases by requiring that $q^o_{ab}$ be the unit
3-sphere or unit 3-hyperboloid metric, in the $k=0$ case there is
no natural way to select a unique $q^o_{ab}$ whence the scale
factor $a(t)$ by itself does not have a direct physical meaning.
This feature introduces certain subtleties in the Hamiltonian
framework in the $k=0$ case. The purpose of this appendix is to
summarize them and correct certain misconceptions that have
permeated in some of the LQG literature.

The Hamiltonian framework has more information than that contained
in the equations of motion. In particular the symplectic structure
can be regarded as the imprint left by the quantum theory on the
classical framework. To define the correct symplectic structure
and Hamiltonian in the non-compact, homogeneous context, one has
to introduce a fiducial cell $\mathcal{V}$ and restrict all
integrations to it. In the $k= -1$ case, one can fix $\mathcal{V}$
to have unit volume with respect to the unit hyperboloid metric.
In the $k=0$ case, such a simple prescription is not available.
However, once $\mathcal{V}$ is chosen, its presence can be used to
make the canonical variables, the symplectic structure and the
Hamiltonian insensitive to the choice of the fiducial metric
$q^o_{ab}$ \cite{abl}. This was the procedure followed in
\cite{aps2} as well as in the present paper. For each choice of
$\mathcal{V}$ we obtain a physical theory. In that theory, one can
make well-defined predictions. However, since in the $k=0$ case
there is no \emph{physical} procedure to fix the fiducial cell, we
now have a new `gauge freedom' $\mathcal{V} \rightarrow \lambda^3
\mathcal{V}$. Under this rescaling the theory changes. Only those
quantities which are insensitive to this rescaling can be regarded
as physical.

Recall that in the quantum theory $\hat{V} \equiv |\hat{p}|^{3/2}$
is the volume operator associated with the fiducial cell
$\mathcal{V}$. Thus, $|\mu\rangle$ (respectively $|v\rangle$) is
an eigenstate of quantum geometry in which the volume of
$\mathcal{V}$ is $ |p|^{3/2} \equiv (8\pi\gamma/6)^{3/2}\,
|\mu|^{3/2}\, \lp^3$ (respectively
$(8\pi\gamma/6)^{3/2}\,(3\sqrt{3\sqrt{3}}/2\sqrt{2})\, v\, \lp^3$).
Now, in the quantum theory based on the cell $\mathcal{V}$,
$|1/p|^{3/2}$ ceases to be a good approximation to the eigenvalues
of the inverse volume operator $\widehat{1/V}$ for $\mu<
\mu_\star$, for a well-defined real number $\mu_\star$ (which
makes no reference to the size fiducial cell $\mathcal{V}$).
Therefore one sometimes finds statements in the LQC literature to
the effect that the quantum geometry effects become important when
$\mu < \mu_\star$. However, physically this would mean that the
quantum geometry effects become important when the physical volume
of $\mathcal{V}$ is less that $|p_\star|^{3/2}$. Clearly, since
the fiducial cell cannot be fixed unambiguously in the $k=0$ case,
unlike in the $k=\pm 1$ cases, this tempting suggestion has no
invariant meaning. Put differently, from a space-time perspective,
one can begin with a quadruplet $(q_{ab}, \phi;\, \dot{q}_{ab},
\dot{\phi})$ representing the initial data at late times for an
expanding universe, consider a semi-classical quantum state which
is sharply peaked about this configuration and ask when, in the
backward quantum evolution, there are significant departures from
the classical trajectory. The answer to this physical question
cannot depend on the choice of the fiducial cell used in the
construction of quantum theory. Therefore, in the $k=0$ case, the
answer that this happens at $p=p_\star$ is not viable. Similarly,
in the analysis of the quantum bounce, the value of $p_{\rm crit}$
(or $v_{\rm crit}$) at which the bounce occurs has only
mathematical meaning within the theory based on a specific
$\mathcal{V}$. From a space-time perspective, this is not
surprising: it is well-known that the value of the scale factor at
which something specific happens is tied to the choice of fiducial
metric and therefore has no invariant meaning.

On the other hand, in the classical theory the matter density
$\rho$ has direct physical meaning also in the $k=0$ case from
both the space-time and the phase space perspectives. In the
space-time perspective, it is given by $\rho = (\dot\phi)^2/2$ and
makes no reference to the fiducial metric. From the phase space
perspective it is given by $\rho = p_\phi^2/2|p|^3$ and is
insensitive to the choice of $\mathcal{V}$ because under the
rescaling $\mathcal{V} \rightarrow \lambda^3 \mathcal{V}$,\,
$p_\phi \rightarrow \lambda^3 p_\phi$ and $|p| \rightarrow
\lambda^2 |p|$. Therefore it is physically meaningful to ask if
there is a critical value of density $\rcr$ at which the classical
theory becomes inadequate. Numerical simulations show that the LQC
dynamics used in the literature yields $\rcr = k|p_{\rm crit}| =
k'/(p_\phi)$, where $k$ and $k'$ are constants that do not depend
on $\mathcal{V}$ and $p_{\rm crit}$ refers to the value of $p$ of
at the bounce point. The physical meaning of $|p|$ and $p_\phi
\equiv \dot{\phi}\, |p|^{3/2}$ depends on the choice of
$\mathcal{V}$. Since $\rho$ itself is gauge invariant, i.e., is
independent of the choice of the fiducial metric $q^o_{ab}$ or the
fiducial cell $\mathcal{V}$, there is a conceptual mismatch.%
\footnote{Furthermore, even if one works with the theory obtained
by just fixing a fiducial cell from the beginning, for states with
$p_\phi \gg \hbar$ (in the c=G=1 units) which are favored by
semi-classical considerations, $\rcr$ can be quite low, violating
the expectation that classical general relativity should be valid
at low densities. It is this second point that was emphasized in
\cite{aps2} and the main body of this paper. The conceptual
mismatch is a related but separate drawback of that quantum
dynamics.}
By contrast, the quantum dynamics presented in this paper leads to
$\rcr \approx 0.82 \rho_{\rm Pl}$ for any state which is
semi-classical at late times; in particular, $\rcr$ makes no
reference to the fiducial cell $\mathcal{V}$.

To summarize, whether one works with geometrodynamics or path
integrals or LQG, the introduction of a fiducial cell
$\mathcal{V}$ is essential in quantization of spatially
non-compact homogeneous models. $\mathcal{V}$ can be used to
remove the reference to the fiducial metric $q^o_{ab}$ from the
theory. However, its presence introduces a new `gauge' freedom and
one has to exercise due care in the physical interpretation of the
theory. If we have two theories, the first based on a fiducial
cell $\mathcal{V}_1$ and the second on $\mathcal{V}_2= \lambda^3
\mathcal{V}_1$, the state $|\mu\rangle$ (or $|v\rangle$) of the
first theory is physically the same as the state
$|\lambda^2\mu\rangle$ (respectively $|\lambda^3 v\rangle$) of the
second theory. Using this equivalence, one has to ensure that the
final physical results are independent of the choice of the
initial fiducial cell.

We will conclude with the discussion of a subtlety. Fix a
fiducial cell $\mathcal{V}$ and consider the resulting quantum
theory. Then, for $v<v_\star$ the inverse volume corrections
become important and the functional form of $\rho (p_\phi, p)$ is significantly
different from the classical relation $\rho = p_\phi^2/2p^3$. Therefore, if
a trajectory were to enter this region, the quantum corrected
$\rho$ along it does not increase and may never reach the critical
value $\rcr$. Then the argument for a quantum bounce would break
down. But a detailed analytical and numerical examination shows
that this could happen only if the assumption $p_\phi \gg \hbar$
(in c=G=1 units) is violated. In a state which is peaked at such a
small $p_\phi$, the uncertainty in the Dirac observables becomes
comparable to the values of the observables themselves, whence the
state can not be regarded as semi-classical. Such a state falls
outside both the detailed quantum evolution studied in the main
body of the paper and the realm of effective equations developed
in Appendix \ref{a2.1}.

\end{appendix}


\begin{thebibliography}{99}

\bibitem{aps2} A.~Ashtekar, T.~Pawlowski and P.~Singh, {Quantum nature
of the big bang: An analytical and numerical investigation I},
Phys. Rev. {\bf D}, at press, \texttt{arXiv:gr-qc/0604013}.

\bibitem{aps1} A.~Ashtekar, T.~Pawlowski and P.~Singh, {Quantum nature
of the big bang}, Phys. Rev. Lett. \textbf{96}, 141301 (2006),
 \texttt{arXiv:gr-qc/0602086}.

\bibitem{mb1} M.~Bojowald, {Absence of singularity in loop
quantum cosmology}, Phys. Rev. Lett. \textbf{86}, 5227-5230
(2001), \texttt{arXiv:gr-qc/0102069}, {Isotropic loop quantum
cosmology}, Class. Quantum. Grav. \textbf{19}, 2717-2741 (2002),
\texttt{arXiv:gr-qc/0202077}.

\bibitem{abl} A.~Ashtekar, M.~Bojowald, J.~Lewandowski,
{Mathematical structure of loop quantum cosmology}, Adv. Theo.
Math. Phys. \textbf{7}, 233-268 (2003), \texttt{gr-qc/0304074}.

\bibitem{mbrev} M.~Bojowald, {Loop quantum cosmology}, Liv.
Rev. Rel. \textbf{8}, 11 (2005), \texttt{arXiv:gr-qc/0601085}.

\bibitem{dh2} K. Banerjee, G. Date, {Discreteness corrections to the
effective Hamiltonian of isotropic loop quantum cosmology}, Class.
Quant. Grav. {\bf 22} (2005) 2017, \texttt{arXiv:gr-qc/0501102};\\
K. Noui, A. Perez, K. Vandersloot, {On the Physical Hilbert Space
of Loop Quantum Cosmology}, Phys. Rev. D {\bf 71} (2005) 044025,
\texttt{gr-qc/0411039}.


\bibitem{alrev} A.~Ashtekar and J.~Lewandowski, {Background
independent quantum gravity: A status report}, Class. Quant. Grav.
{\bf 21}, R53-R152 (2004), \texttt{arXiv:gr-qc/0404018}.

\bibitem{crbook} C.~Rovelli {\em Quantum Gravity}, (CUP,
Cambridge, 2004).

\bibitem{ttbook} T.~Thiemann, {\em Introduction to Modern Canonical
Quantum General Relativity} (CUP, Cambridge, at press).

\bibitem{almmt} A.~Ashtekar, J.~Lewandowski, D.~Marolf,
J.~Mour\~ao and T.~Thiemann, Quantization of diffeomorphism
invariant theories of connections with local degrees of freedom
{Jour. Math. Phys.} \textbf{36} 6456--6493 (1995),
\texttt{arXiv:gr-qc/9504018}.

\bibitem{rs} C.~Rovelli and L.~Smolin Discreteness of area and volume
in quantum gravity  \textit{Nucl. Phys.} \textbf{B442}  593--622
(1995); Erratum: \textit{Nucl. Phys.} \textbf{B456} 753, (1996)
\texttt{gr-qc/9411005}.

\bibitem{al5} A.~Ashtekar and J.~Lewandowski, Quantum theory of
geometry I: Area operators, {Class. Quant. Grav.} \textbf{14}
A55--A81 (1997), \texttt{arXiv:gr-qc/9602046}.

\bibitem{aaparis} A.~Ashtekar, Geometry, gravity and the quantum,
in the \emph{Proceedings of the Einstein Century Conference},
edited by J-M Alimi et al, (AIP, New York, in Press),
\texttt{arXiv:gr-qc/0605011}.

\bibitem{afw} A.~Ashtekar, S.~Fairhurst and J.~Willis, {Quantum
gravity, shadow states, and quantum mechanics}, Class.\ Quantum\
Grav. \textbf{20}, 1031-1062 (2003), \texttt{arXiv:gr-qc/0207106}.



\bibitem{tt} Thiemann, T. Anomaly-free formulation of
non-perturbative, four-dimensional Lorentzian quantum gravity,
\textit{Phys. Lett.} \textbf{B380}, 257--264 (1998),
\texttt{gr-qc/9606088}, \\
Quantum spin dynamics (QSD), {Class. Quant. Grav.} \textbf{15}
839--873 (1998), \texttt{gr-qc/9606089}, \\ QSD V : Quantum
gravity as the natural regulator of matter quantum field theories,
{Class. Quant. Grav.} \textbf{15}, 1281--1314 (1998),
\texttt{gr-qc/9705019}.

\bibitem{boj-pc} M. Bojowald, Personal Communication (2006).

\bibitem{nmjw} N.~M.~J~Woodhouse, \emph{Geometric Quantization},
(Oxford UP, Oxford, 1997).

\bibitem{mb-amb} M.~Bojowald, Quantization ambiguities in isotropic
quantum geometry, Class. Quant. Grav. {\bf 19}, 5113 (2002),
\texttt{gr-qc/0206053}.

\bibitem{kv} K.~Vandersloot, On the Hamiltonian constraint of loop
quantum cosmology, Phys. Rev. D\textbf{71} 103506 (2005),
\texttt{gr-qc/0502082};\\
Ph.D. Dissertation, submitted to The Pennsylvania State University
(2006).

\bibitem{ap} A.~Perez, {On the regularization ambiguities in loop
quantum gravity}, Phys.Rev. D {\bf 73} (2006) 044007,
\texttt{arXiv:gr-qc/0509118}.




\bibitem{wk} W.~Kaminski, personal communication to AA (November
2005).



\bibitem{ag} A.~Ashtekar and R.~Geroch, Quantum theory of
gravitation, Rep. Prog. Phys. \textbf{37}, 1211-1256 (1974).

\bibitem{ck} C.~Kiefer, Wave packets in minisuperspace, Phys. Rev.
Rev. \textbf{D38}, 1761-1772 (1988).

\bibitem{aabook} A.~Ashtekar, \textit{Lectures on non-perturbative
canonical gravity}, Notes prepared in collaboration with R. S.
Tate (World Scientific, Singapore, 1991), Chapter 10.

\bibitem{at} A.~Ashtekar and R.~S.~Tate, An algebraic extension
of Dirac quantization: Examples, {Jour. Math. Phys.} \textbf{35}
6434--6470 (1994).




\bibitem{dm} D.~Marolf, Refined algebraic quantization:
Systems with a single constraint \texttt{arXives:gr-qc/9508015};\\
{Quantum observables and recollapsing dynamics}, Class. Quant.
Grav. {\bf 12} (1995) 1199-1220
\texttt{arXiv:gr-qc:9404053}\\
{Observables and a Hilbert space for Bianchi IX}, Class. Quant.
Grav. {\bf 12} (1995) 1441-1454
\texttt{arXiv:gr-qc:9409049}\\
{Almost ideal clocks in quantum cosmology: A brief derivation of
time}, Class. Quant. Grav. {\bf 12} (1995) 2469-2486
\texttt{arXiv:gr-qc:9412016}.

\bibitem{hm2} J. B.~Hartle and D.~Marolf, {Comparing formulations of
generalized quantum mechanics for reparametrization-invariant
systems}, Phys. Rev. \textbf{D56}, 6247-6257 (1997)
\texttt{arXiv:gr-qc:9703021}.



\bibitem{singh:2006a} P. Singh, Loop cosmological dynamics and
dualities with Randall-Sundrum braneworlds, Phys. Rev. {\bf D73}
063508 (2006), \texttt{arXiv:gr-qc/0603043}.

\bibitem{as}A.~Ashtekar and T.~Schilling, Geometrical formulation of
quantum mechanics, In: \textit{On Einstein's path}, A.~Harvery, ed
(Springer-Verlag, New York, 1998); \texttt{arXiv:gr-qc/9706069}.
T.~Schilling, Geometry of quantum mechanics, Ph.D. Dissertation,
Penn State (1996),
http://cgpg.gravity.psu.edu/archives/thesis/index.shtml

\bibitem{jw} J.~Willis, {On the low energy ramifications and a
mathematical extension of loop quantum gravity}, Ph.D.
Dissertation, The Pennsylvania State University (2004);
A.~Ashtekar, M.~Bojowald and J.~Willis, {Corrections to Friedmann
equations induced by quantum geometry}, IGPG preprint (2004).

\bibitem{vt} V. Taveras, IGPG preprint (2006).

\bibitem{imp_pheno} P.~Singh, K.~Vandersloot,
G.~V.~Vereshchagin, Non-Singular Bouncing Universes in Loop
Quantum Cosmology, Phys. Rev. \textbf{D74}  043510 (2006); \texttt{arXiv:gr-qc/0606032}.

\bibitem{mr} M.~Reuter and F.~Saueressig, From big bang to
asymptotic de Sitter: Complete cosmologies in a quantum gravity
framework, JCAP 0509, 012 (2005), \texttt{hep-th/0507167}.


\end{thebibliography}
\end{document}